\documentclass[ALICE,manyauthors]{cernphprep}

\usepackage[comma,square,numbers,sort&compress]{natbib}
\usepackage{hyperref}
\usepackage{lineno}
\usepackage{float}
\usepackage{color}
\usepackage{booktabs}
\usepackage{multirow}
\usepackage{soul}

\usepackage{array}
\usepackage{makecell}

\newcommand{\pt}{\ensuremath{p_{\rm T}}\xspace}
\newcommand{\mpt}{\ensuremath{\langle p_{\rm T} \rangle}\xspace}

\begin{document}%

\begin{titlepage}
\PHyear{2019}
\PHnumber{094}      
\PHdate{7 May}  
%

\title{Charged-particle production as a function of multiplicity and transverse spherocity in pp collisions at $\mathbf{ \sqrt{s} =5.02}$ and 13\,TeV }
\ShortTitle{Particle production as a function of multiplicity and spherocity in pp collisions}   

\Collaboration{ALICE Collaboration\thanks{See Appendix~\ref{app:collab} for the list of collaboration members}}
\ShortAuthor{ALICE Collaboration} 

\begin{abstract}
We present a study of the inclusive charged-particle transverse momentum ($p_{\rm T}$) spectra as a function of charged-particle multiplicity density at mid-pseudorapidity, ${\rm d}N_{\rm ch}/{\rm d}\eta$, in pp collisions at $\sqrt{s}=5.02$ and 13\,TeV covering the kinematic range $|\eta|<0.8$ and $0.15<\pt<20$\,GeV/$c$. The results are presented for events with at least one charged particle in $|\eta|<1$ (INEL$\,>0$). The $p_{\rm T}$ spectra are reported for two multiplicity estimators covering different pseudorapidity regions. The $p_{\rm T}$ spectra normalized to that for INEL$\,>0$ show little energy dependence. Moreover, the high-$p_{\rm T}$ yields of charged particles increase faster than the charged-particle multiplicity density.
The average $\it{p}_{\rm T}$ as a function of multiplicity and transverse spherocity is reported for pp collisions at $\sqrt{s}=13$\,TeV. For low- (high-) spherocity events, corresponding to jet-like (isotropic) events, the average $p_{\rm T}$ is higher (smaller) than that measured in INEL$\,>0$ pp collisions. Within uncertainties, the functional form of $\langle p_{\rm T} \rangle(N_{\rm ch})$ is not affected by the spherocity selection. While EPOS~LHC gives a good description of many features of data, PYTHIA overestimates the average $p_{\rm T}$ in jet-like events.

\end{abstract}
\end{titlepage}
\setcounter{page}{2}

\section{Introduction}
Proton-proton collisions at the Large Hadron Collider (LHC) energies have unveiled features very similar to the ones observed in heavy-ion collisions~\cite{Loizides:2016tew}. The previous consensus of the heavy-ion community was that the partonic system created in nuclear collisions needs a large volume to thermalize and to lead to phenomena like collective flow.  However, radial~\cite{Abelev:2013haa, Adam:2016dau,Acharya:2018orn} and anisotropic flow~\cite{Khachatryan:2016txc}, as well as strangeness enhancement~\cite{ALICE:2017jyt}, are also observed in pp and p-A collisions when they are studied as a function of event multiplicity. Surprisingly, with the same level of precision, microscopic and macroscopic approaches describe qualitatively well the observed features in pp collisions. While macroscopic models incorporate  hydrodynamical evolution of the system~\cite{Werner:2013ipa}, the others include overlapping strings~\cite{Bierlich:2014xba}, string percolation~\cite{Bautista:2015kwa}, multi-parton interactions and color reconnection~\cite{Sjostrand:2014zea,PhysRevLett.111.042001}. The multiphase transport model~\cite{PhysRevC.92.054903}, as well as the fragmentation of saturated gluon states~\cite{Schlichting:2016sqo,Schenke:2016lrs}, is able to describe some features of data. 

The inclusive transverse momentum (\pt) spectrum of charged particles carries information of the dynamics of soft and hard interactions. On one hand, the high-\pt ($\pt>10$\,GeV/$c$) particle production is quantitatively well described by perturbative QCD (pQCD) calculations; on the other hand, the understanding of particle production at low-\pt has to resort to phenomenological QCD inspired models. Most of the new effects discovered in pp collisions have been unveiled in the low- ($\pt<2$\,GeV/$c$) and intermediate- ($2 \leq \pt<10$\,GeV/$c$) \pt domains~\cite{Abelev:2013haa, Adam:2016dau,Acharya:2018orn,Khachatryan:2016txc,ALICE:2017jyt}. The present paper reports a novel multi-differential analysis aimed at understanding charged-particle production associated to partonic scatterings with large momentum transfer and their possible correlations with soft particle production.

The transverse momentum distributions are reported for two multiplicity estimators which cover different pseudorapidity regions. The estimators are based on either the total charge deposited in the forward detector (covering the pseudorapidity regions $2.8<\eta<5.1$ and $-3.7<\eta<-1.7$) or on the number of tracks in the pseudorapidity region $|\eta|<0.8$.   The forward multiplicity estimator is commonly used by the ALICE collaboration to minimize the possible autocorrelations induced by the use of the mid-pseudorapidity estimator. One such examples is the ``fragmentation bias''~\cite{Abelev:2013sqa}, which is the correlation between jet fragments and event multiplicity arising when the particle's \pt and event multiplicity are both measured within the same pseudorapidity interval~\cite{Ortiz:2016kpz}. For each estimator, we defined different multiplicity classes based on either the number of tracks at mid-pseudorapidity ($|\eta|<0.8$) or the signal in the forward detectors. It is worth mentioning that a similar study has been performed by ALICE using p-Pb data; the results showed different modifications of the spectral shapes depending on  the multiplicity estimators which were used~\cite{Adam:2014qja}. To disentangle the energy and multiplicity dependence, for a given multiplicity class, the \pt distributions are measured for pp collisions at $\sqrt{s}=5.02$ and 13\,TeV. Particle production from intermediate to high \pt ($>4$\,GeV/$c$) is studied by fitting a power-law function to the invariant yield,  and studying the multiplicity and energy dependence of the exponent. This has been proposed in Ref.~\cite{Arleo:2009ch} as a way to characterize the high-\pt tails of different systems and energies in a convenient way that may make the comparison for the different systems more straightforward. 

Finally, we explore a new approach, which has been proposed to study multi-parton interaction effects in pp collisions. Transverse spherocity~\cite{Cuautle:2015kra}, hereinafter referred to as spherocity, has been proven to be a valuable tool to discriminate between jet-like and isotropic events~\cite{Ortiz:2017jho} associated with an underlying event activity which is either suppressed or enhanced. The previous measurement of average transverse momentum of inclusive charged particles as a function of event multiplicity~\cite{Abelev:2013bla} is now explored adding a new dimension: the event shape characterized by spherocity.  The aim of this study is to investigate the importance of jets in high-multiplicity pp collisions and their contribution to charged-particle production at low \pt.

The paper is organized as follows: Section~\ref{s:1} describes the run conditions during the data taking and the main detectors used in the present analysis. Section~\ref{s:2} outlines the analysis details for the event and track selection, as well as the definitions of the different event classes. The correction procedures and the estimation of the systematic uncertainties are summarized in Sections~\ref{s:3} and~\ref{s:4}, respectively. Results and discussions are presented in Section~\ref{s:5}. Finally, our summary and conclusions are reported in Section~\ref{s:6}.

\section{The ALICE apparatus}\label{s:1}

The main detectors used in the present work are the Inner Tracking System (ITS), the Time Projection Chamber (TPC) and the V0 detector.  The ITS and TPC detectors are both used for primary vertex and track reconstruction. The V0  detector is used for triggering and for background rejection.  More details concerning the full ALICE detector system can be found in Ref.~\cite{Abelev:2014ffa}.

The central barrel covers the pseudorapidity region $|\eta|<0.8$ for full-length tracks.  The main central-barrel tracking devices are the ITS and the TPC, which are located inside a solenoid magnet providing a 0.5\,T magnetic field allowing the tracking of particles from 0.15\,GeV/$c$.  The ITS is composed of six cylindrical layers of high-resolution silicon tracking detectors.  The innermost layers consist of two arrays of hybrid Silicon Pixel Detectors (SPD) located at an average radial distance ($r$) of 3.9 and 7.6\,cm from the beam axis and covering $|\eta|<2$ and $|\eta|<1.4$, respectively.  The SPD is also used to reconstruct tracklets, which are track segments built using the position of  the reconstructed  primary  vertex  and  two  hits,  one  on  each  SPD  layer. The number of tracklets gives an excellent estimate of the charged-particle multiplicity at mid-pseudorapidity ($N_{\rm ch}$). The outer layers of the ITS are composed of silicon strip and drift detectors, with the outermost layer sitting at $r=43$\,cm.  The TPC is a large cylindrical drift detector of radial and longitudinal size of about $85<r<250$\,cm and $-250<z<250$\,cm, respectively.  It is segmented in radial ``pad rows'', providing up to 159 tracking points.  The measurement of charged particles is based on ``global tracks'', reconstructed using the combined ITS and TPC information.  The V0 detector consists of two forward scintillator arrays (V0-A and V0-C) employed for triggering, background suppression, and event-class determination. They are placed on either side of the interaction region at $z=3.3$\,m and $z=-0.9$\,m,  covering the pseudorapidity regions $2.8<\eta<5.1$ and $-3.7<\eta<-1.7$, respectively.

The data were collected using a minimum-bias trigger which required coincident signals in both V0-A and V0-C detectors .  The events were recorded in coincidence with signals from two beam pick-up counters each positioned on either side of the interaction region to tag the arrival of proton bunches from both directions. Control triggers taken for various combinations of beam and empty buckets were used to measure beam-induced and accidental backgrounds.  The contamination from background events was removed offline by using the timing information from the V0 detector, which has a time resolution better than 1\,ns. Background events were also rejected by exploiting the correlation between the number of clusters of pixel hits and the number of tracklets in the SPD.

\section{Analysis}\label{s:2}

The results presented here were obtained from the analysis of about 105 and 60 million minimum-bias pp events at $\sqrt{s}=5.02$ and 13\,TeV, respectively. The interaction probability per single bunch crossing ranges between 2\% and 14\% for pp collisions at\,13 TeV and from 0.3\% to 6\% for pp collisions at 5.02\,TeV. The measurements have been obtained for events having at least one charged particle produced in the pseudorapidity interval $|\eta|<1$ (INEL$\,>0$). For the analysis, the events were furthermore required to have a reconstructed vertex located within  $|z|<10$\,cm, where $z$ is the position of the vertex along the beam axis, and  $z=0$\,cm corresponds to the nominal center of the detector~\cite{Abelev:2014ffa}. Events containing more than one distinct vertex were tagged as pileup and discarded from the analysis.  The systematic uncertainty associated to pileup is between 3\,--\,4\% and is not the dominant source of uncertainty for the \pt spectra reported here. The corrections are calculated using Monte Carlo events from the  PYTHIA~6~\cite{Sjostrand:2006za} (tune Perugia~2011~\cite{Skands:2010ak})  event generator with particle transport performed via a GEANT~3~\cite{Brun:1994aa} simulation of the ALICE detector.

Only primary charged particles in the kinematic range $|\eta|<0.8$ and $0.15<\pt<20$\,GeV/$c$ are considered in the transverse momentum analysis. A primary charged particle is defined as a charged particle with a mean proper lifetime $\tau$ larger than 1\,cm$/c$, which is either produced directly in the interaction or from decays of particles with $\tau$ smaller than 1\,cm$/c$, excluding particles produced in interactions with the detector material~\cite{ALICE-PUBLIC-2017-005}.

\paragraph{Transverse momentum distributions.} The measurement of the \pt spectra follows the standard procedure already employed in several ALICE publications~\cite{Abelev:2013ala,Adam:2015pza,Acharya:2018qsh}. Tracks reconstructed using the information from the ITS and TPC detectors are used. The track selection criteria have been optimised for best track quality and minimal contamination from secondary particles. Tracks are required to have at least two hits in the ITS detector, of which at least one is in either of the two innermost SPD layers. The geometrical track length $L$ (in cm) is calculated in the TPC readout plane, excluding the information from the pads at the sector boundaries ($\sim$3\,cm from the sector edges). The number of crossed TPC rows has to be larger than 0.85$L$. The number of TPC clusters has to be larger than 0.7$L$. The fit quality for the ITS and TPC track points must satisfy $\chi^{2}_{\rm ITS}/N_{\rm hits}<36$ and $\chi^{2}_{\rm TPC}/N_{\rm clusters}<4$, respectively, where $N_{\rm hits}$ and $N_{\rm clusters}$ are the numbers of hits in the ITS and the number of clusters in the TPC, respectively. Tracking information from  the combined ITS and TPC track reconstruction algorithm  is compared to that derived only from the TPC and constrained by the interaction  vertex point. Then, the quantity $\chi^{2}_{\rm TPC-ITS}$ is derived as described in Ref.~\cite{Abelev:2012hxa}. Only tracks with  $\chi^{2}_{\rm TPC-ITS}<36$ are included in the analysis  in order to improve the purity of primary track reconstruction at high \pt. Tracks are rejected if their distance of closest approach to the reconstructed vertex in longitudinal and radial direction, $d_{z}$ and $d_{xy}$, respectively, satisfies $d_{z}>2$\,cm or $d_{xy}>0.018$\,cm\,$+$\,$0.035$\,cm\,$\times \pt^{-1.01}$, with \pt in GeV/$c$.

\paragraph{Multiplicity estimators.} In order to study the multiplicity dependence of the inclusive charged particle \pt distributions, the INEL$\,>0$ sample is divided into event classes based on the total charge deposited in the V0 detector (V0M amplitude) and on the number of SPD tracklets ($N_{\rm SPD\,tracklets}$) in the pseudorapidity region $|\eta|<0.8$. The event classes used in the analysis and the corresponding mid-pseudorapidity charged particle densities are summarized in Tables~\ref{tab:1a}~and~\ref{tab:1b}. The average charged-particle multiplicity densities for INEL$\,>0$ collisions and for the multiplicity classes are obtained by integrating the corresponding fully corrected \pt spectra (measured using ITS and TPC information). To this end, the \pt spectra were extrapolated to $\pt=0$ with a Hagedorn function~\cite{Hagedorn:1983}. Different functions were used and the differences with respect to the reference values were considered in the systematic uncertainties.  For INEL$\,>0$ pp collisions at $\sqrt{s}=5.02$\,TeV the mid-pseudorapidity ($|\eta|<0.8$) charged-particle density is $\langle {\rm d}N_{\rm ch}/{\rm d}\eta \rangle=5.91\pm0.45$, while for $\sqrt{s}=13$\,TeV the corresponding value is $7.60\pm 0.50$. The comparison of results obtained with these estimators allows to understand potential biases from measuring the multiplicity and \pt distributions in overlapping $\eta$ regions. 

\begin{table}[h!]
\centering

   \caption{V0M event multiplicity classes, their corresponding experimental definition and their corresponding $\langle {\rm d}N_{\rm ch}/{\rm d}\eta \rangle$ in $|\eta|<0.8$. The uncertainties are the quadratic sum of statistical and systematic contributions. Statistical uncertainties are negligible compared to the systematic ones.}
\begin{tabular}{ p{3cm} p{2cm} p{2cm} p{2cm}  p{2cm} p{2cm}  }
 \hline
 \multicolumn{6}{c}{pp collisions at $\sqrt{s}=13$\,TeV} \\
 \hline
 Class name & I & II & III  &  IV & V     \\
 \hline
    V0M percentile              &  0\,--\,1\%   &   1\,--\,5\%    &  5\,--\,10\%  &   10\,--\,15\%    & 15\,--\,20\%   \\
    $\langle {\rm d}N_{\rm ch}/{\rm d}\eta \rangle$ &  26.6$\pm$1.1   &    20.5$\pm$0.8   &  16.7$\pm$0.7  &  14.3$\pm$0.6     & 12.6$\pm$0.5   \\
 \hline
 \hline
  Class name & VI & VII & VIII  &  IX & X     \\
 \hline  
    V0M percentile             &  20\,--\,30\%   &   30\,--\,40\%    &  40\,--\,50\%  &   50\,--\,70\%    & 70\,--\,100\%   \\
    $\langle {\rm d}N_{\rm ch}/{\rm d}\eta \rangle$ &  10.6$\pm$0.5   &    8.46$\pm$0.40   &  6.82$\pm$0.34  &  4.94$\pm$0.28     & 2.54$\pm$0.26   \\ 
 \hline
  \hline
 \multicolumn{6}{c}{pp collisions at $\sqrt{s}=5.02$\,TeV} \\
 \hline
 Class name & I & II & III  &  IV & V     \\
 \hline
    V0M percentile              &  0\,--\,1\%   &   1\,--\,5\%    &  5\,--\,10\%  &   10\,--\,15\%    & 15\,--\,20\%   \\
    $\langle {\rm d}N_{\rm ch}/{\rm d}\eta \rangle$ &  19.2$\pm$0.9   &    15.1$\pm$0.7   &  12.4$\pm$0.6  &  10.7$\pm$0.5     & 9.47$\pm$0.47   \\
 \hline
 \hline
  Class name & VI & VII & VIII  &  IX & X     \\
 \hline  
    V0M percentile              &  20\,--\,30\%   &   30\,--\,40\%    &  40\,--\,50\%  &   50\,--\,70\%    & 70\,--\,100\%   \\
    $\langle {\rm d}N_{\rm ch}/{\rm d}\eta \rangle$ &  8.04$\pm$0.42   &    6.56$\pm$0.37   &  5.39$\pm$0.32  &  4.05$\pm$0.27     & 2.27$\pm$0.27   \\ 
 \hline
\end{tabular}
   \label{tab:1a}
\end{table}

\begin{table}[h!]
\centering
   \caption{Event multiplicity classes based on the number of tracklets ($N_{\rm SPD\,tracklets}$) within $|\eta|<0.8$, their corresponding experimental definition and their corresponding $\langle {\rm d}N_{\rm ch}/{\rm d}\eta \rangle$ in $|\eta|<0.8$. The uncertainties are the quadratic sum of statistical and systematic contributions.}
\begin{tabular}{ p{3cm} p{2cm} p{2cm} p{2cm}  p{2cm}p{2cm}  }
 \hline
 \multicolumn{6}{c}{pp collisions at $\sqrt{s}=13$\,TeV} \\
 \hline
 \thead{Class name \\ (percentile)}& \thead{ I$^{\prime}$ \\ 0--0.006\% } & \thead{II$^{\prime}$ \\ 0.006--0.058\% } &  \thead{III$^{\prime}$ \\ 0.058--0.177\%}  &  \thead{IV$^{\prime}$ \\ 0.177--0.513\%}   &  \thead{V$^{\prime}$ \\0.513--1.419\%}  \\
 \hline
    $N_{\rm SPD\,tracklets}$ & $\geq$ 51 & 41\,--\,50 & 36\,--\,40 & 31\,--\,35 & 26\,--\,30  \\
    $\langle {\rm d}N_{\rm ch}/{\rm d}\eta \rangle$ &  54.1$\pm$2.7   &    44.6$\pm$2.2   &  38.9$\pm$1.9  &  34.1$\pm$1.7     & 29.3$\pm$1.5   \\
 \hline
 \hline
  \thead{Class name \\ percentile } & \thead{VI$^{\prime}$\\1.419--3.699\%} & \thead{VII$^{\prime}$ \\ 3.699--9.059\%}   &  \thead{VIII$^{\prime}$ \\ 9.059--20.77\%}  &  \thead{IX$^{\prime}$ \\ 20.77--45.25\%}  &  \thead{X$^{\prime}$\\ 45.25--100.0\%}   \\
 \hline  
  $N_{\rm SPD\,tracklets}$ &   21\,--\,25 & 16\,--\,20 & 11\,--\,15 & 6\,--\,10 & 0\,--\,5  \\
    $\langle {\rm d}N_{\rm ch}/{\rm d}\eta \rangle$ &  24.5$\pm$1.3   &    19.5$\pm$1.2   &  14.4$\pm$0.9  &  9.03$\pm$0.58     & 2.91$\pm$0.29   \\ 
 \hline
  \hline
 \multicolumn{6}{c}{pp collisions at $\sqrt{s}=5.02$\,TeV} \\
 \hline
\thead{Class name \\ (percentile)} & - & \thead{ II$^{\prime}$ \\ 0.009--0.088\%}   &  \thead{III$^{\prime}$ \\ 0.088--0.253\%}   &  \thead{IV$^{\prime}$ \\ 0.253--0.700\%}  &  \thead{V$^{\prime}$ \\ 0.700--1.840\%}   \\
 \hline
     $N_{\rm SPD\,tracklets}$  & - & 41\,--\,50 & 36\,--\,40 & 31\,--\,35 & 26\,--\,30  \\
    $\langle {\rm d}N_{\rm ch}/{\rm d}\eta \rangle$ & - & 34.6$\pm$1.8   &    29.9$\pm$1.5   &  26.2$\pm$1.3  &  22.4$\pm$1.1  \\
 \hline
 \hline
  \thead{Class name \\ (percentile) } & \thead{VI$^{\prime}$\\1.840--4.573\%} & \thead{VII$^{\prime}$\\4.57--10.69\%}   &  \thead{VIII$^{\prime}$ \\ 10.69--23.50\%}  &  \thead{IX$^{\prime}$\\23.50--49.48\%}  & \thead{X$^{\prime}$\\49.48--100.0\%}    \\
 \hline  
 $N_{\rm SPD\,tracklets}$ & 21\,--\,25 & 16\,--\,20 & 11\,--\,15 & 6\,--\,10 & 0\,--\,5 \\    
    $\langle {\rm d}N_{\rm ch}/{\rm d}\eta \rangle$ & 18.5$\pm$1.0  &14.6$\pm$0.9   &    10.6$\pm$0.7   &  6.58$\pm$0.43  &  2.21$\pm$0.24     \\ 
 \hline
\end{tabular}
   \label{tab:1b}
\end{table}

\paragraph{Spherocity.} For the data analysis we followed a strategy similar to that already reported in Ref.~\cite{Abelev:2012sk}. Spherocity, $S_{0}$, originally proposed here~\cite{Banfi:2010xy} is defined for a unit vector $\mathbf{ \hat{\rm \mathbf{n}}_{\rm \mathbf{s}} }$ which minimizes the ratio:

\begin{equation}
S_{\rm 0} \equiv \frac{\pi^{2}}{4}  \underset{\bf \hat{n}_{\rm \bf{s}}}{\text{min}}  \left( \frac{\sum_{i}|{\vec p}_{{\rm T},i} \times { \bf \hat{n}_{\rm \bf{s}} }|}{\sum_{i}p_{{\rm T},i}}  \right)^{2},
\end{equation}
where the sum runs over all reconstructed ITS-TPC tracks. At least three tracks are required within $|\eta|<0.8$ and $\pt>0.15$\,GeV/$c$ in order to achieve a good spherocity resolution. The spherocity resolution improves with the track-reconstruction efficiency, therefore the restrictions on the purity of primary charged particles can be relaxed. For spherocity we considered all tracks with at least 50 clusters in the TPC, which satisfy: $d_{\rm xy}<2.4$\,cm, $d_{z}<3.2$\,cm, and $\chi^{2}_{\rm TPC}/N_{\rm clusters}<4$. The exclusion of the ITS requirements guarantees a homogeneous azimuthal track-reconstruction efficiency.

It is worth mentioning some important features of spherocity:
\begin{itemize}
\item The vector products are linear in particle momenta, therefore spherocity is a collinear safe quantity in pQCD.
\item The lower limit of spherocity ($S_{0}\rightarrow0$) corresponds to event topologies where all transverse momentum vectors are (anti)parallel or the sum of the \pt is dominated by a single track.
\item The upper limit of spherocity ($S_{0}\rightarrow1$) corresponds to event topologies where transverse momentum vectors are ``isotropically'' distributed. $S_{0}=1$ can only be reached in the limit of an infinite amount of particles.
\end{itemize}

\begin{figure}[htb]
\centerline{
\includegraphics[width=0.5\columnwidth]{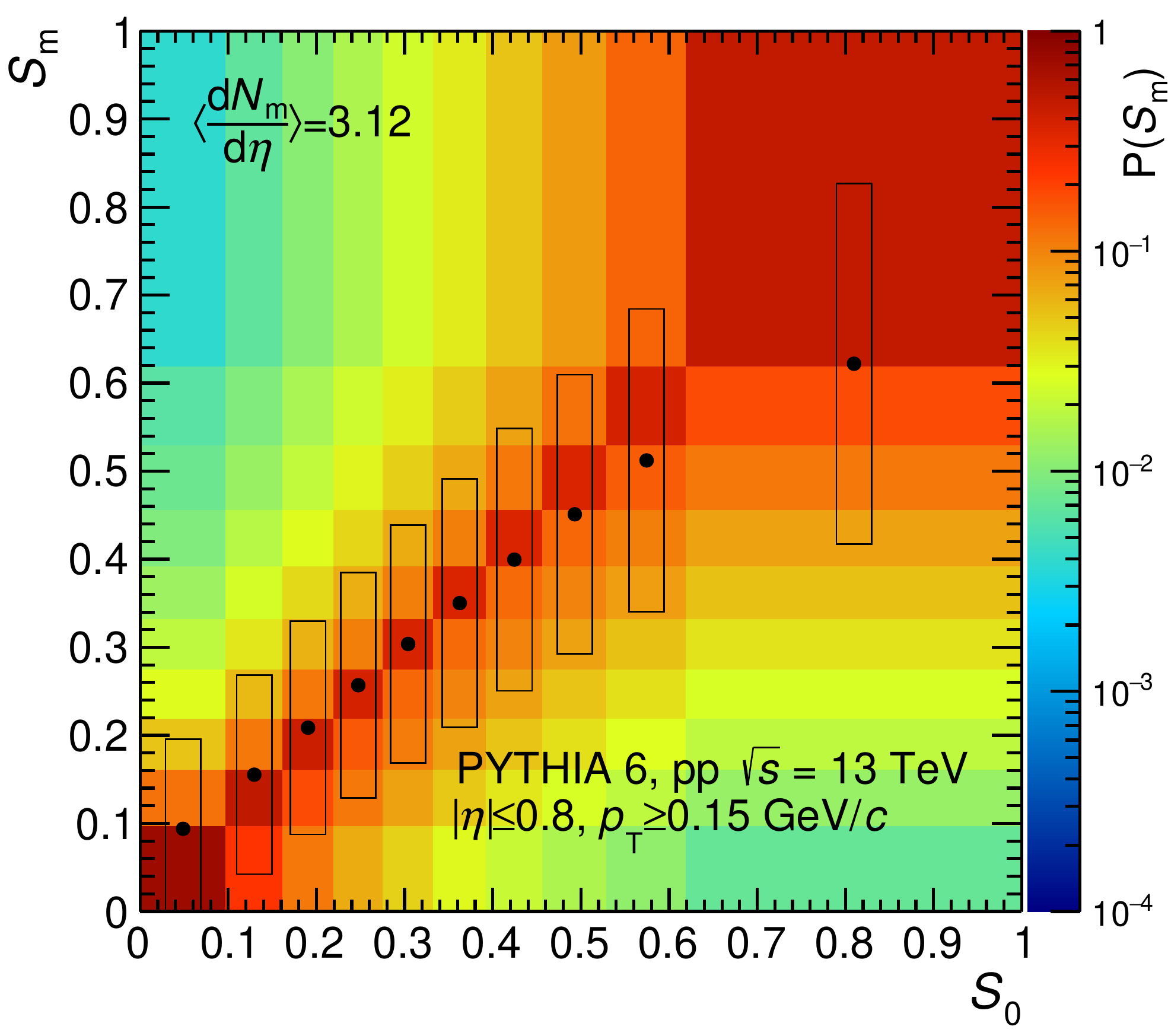}
\includegraphics[width=0.5\columnwidth]{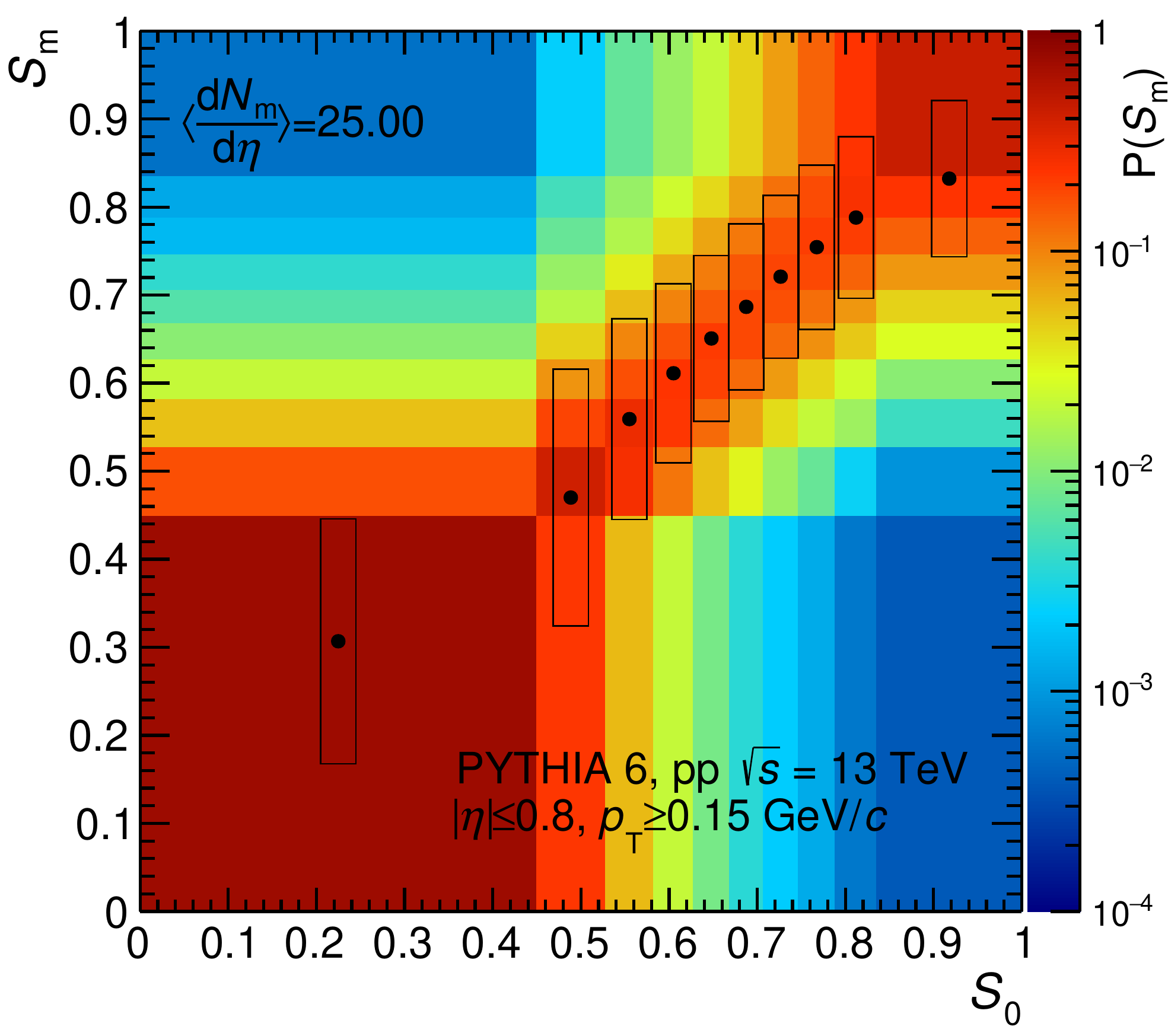}}
\caption{ The detector response for spherocity for two track multiplicity classes: $\langle {\rm d}N_{\rm m}/{\rm d}\eta \rangle=3.12$ (left) and $\langle {\rm d}N_{\rm m}/{\rm d}\eta\rangle=25$ (right). Proton-proton collisions were simulated using PYTHIA~6. The simulations include the particle transport performed via a GEANT~3 simulation of the ALICE detector. The markers (boxes around the points) indicate the average (RMS) of the measured spherocity distributions for each bin of spherocity at generator level. The spherocity binning varies with ${\rm d}N_{\rm m}/{\rm d}\eta$,  because in this way, it allows the analysis of ten event sub-classes of equal size.  The probability that an event with spherocity $S_{0}$ be reconstructed with spherocity $S_{\rm m}$ is represented by ${\rm P}(S_{\rm m})$.}
\label{so:1}
\end{figure}

Since the goal of the present study is to separate jet events from isotropic ones, we study different spherocity classes for a given multiplicity value. The multiplicity is measured by counting the number of tracks within $|\eta|<0.8$. As explained later, we adopted the procedure used in the analysis of average \pt as a function of multiplicity to correct the number of tracks for detector effects~\cite{Abelev:2013bla}. The detector response is represented by a two-dimensional distribution: reconstructed spherocity as a function of generated spherocity, each bin of generated spherocity is normalized to unity. In this representation, the two-dimensional distribution gives the normalized response matrix $R^{\prime}(S_{0},S_{\rm m})$, which contains the probability that an event with spherocity $S_{0}$ is reconstructed with spherocity $S_{\rm m}$. Figure~\ref{so:1} shows the spherocity response matrices for two track multiplicity ($N_{\rm m}$). Tracking efficiency effects on the spherocity resolution are relevant only for low-multiplicity events, therefore, the $S_{0}$ resolution improves with increasing multiplicity. 

In order to study the spherocity dependence of the particle production for a given track multiplicity value, the sample is divided into ten event sub-classes of equal size  (percentiles), based on the measured spherocity distribution. From now on, the most jet-like and isotropic events will be referred to as 0\,--\,10\% and 90\,--\,100\% spherocity event class, respectively. 

It has been reported that the evolution of several observables as a function of center-of-mass energy can be factored out to be due to the changes in charged-particle multiplicities which in turn depend on the energy. For example, the particle production sensitive to the underlying event for different $\sqrt{s}$ exhibits approximate scaling properties connected to changes in $\langle N_{\rm ch} \rangle$~\cite{Ortiz:2017jaz}. Moreover, within uncertainties, the average \pt as a function of multiplicity exhibits a small energy dependence~\cite{Abelev:2013bla}. Therefore, the spherocity dependent average \pt as a function of charged-particle multiplicity is only presented for pp collisions at $\sqrt{s}=13$\,TeV. The physics message is valid for other center-of-mass energies, this was verified using data from pp collisions at $\sqrt{s}=5.02$\,TeV.

\section{Corrections}\label{s:3}

All the measurements presented in this paper are fully corrected for  acceptance and tracking efficiency, contamination from secondary particles, event and signal loss, as well as multiplicity and spherocity resolution. Details of these corrections are presented below.

\subsection{Transverse momentum distributions as a function of particle multiplicity}

The transverse momentum spectrum for a specific event class is obtained by correcting the track yields $N^{\rm rec}$ reconstructed in each $(\Delta\eta,\Delta\pt)$ interval for all detector effects that either influence the event reconstruction or the track reconstruction. The transverse momentum distribution is obtained as follows:
\begin{equation}
\frac{1}{N_{\rm ev}}\frac{{\rm d}^{2}N_{\rm ch}}{{\rm d}\eta{\rm d}\pt} \equiv \frac{N^{\rm rec}(\eta,\pt)  C(\eta,\pt)}{N_{\rm ev}^{\rm rec}  \Delta\eta  \Delta\pt}  \epsilon_{\rm ev.\,class} \epsilon_{\rm vz}.
\end{equation}
The event selection (for a specific event class) and vertex reconstruction efficiencies are represented by $\epsilon_{\rm ev.\,class}$ and $\epsilon_{\rm vz}$, respectively.  The number of events of a given event class is represented by $N_{\rm ev}^{\rm rec}$. For the lowest multiplicity class selected using the V0M amplitude and for  $\sqrt{s}=5.02$\,TeV ($\sqrt{s}=13$\,TeV) they reach 66\% and 95\% (75\% and 95\%), respectively, while for the highest multiplicity class the detector is fully efficient.  The track-level correction factors, $C(\Delta\eta,\Delta\pt)$, are obtained for events which satisfy the selection criteria; they include acceptance, efficiency, purity, and \pt resolution. The estimation of the four terms will be explained in detail in the following.

A data-driven method has been developed to reduce the systematic uncertainty related to incorrect description of the particle composition in Monte Carlo. The tracking efficiency is determined using the re-weighting procedure which is discussed for the first time in Ref.~\cite{Acharya:2018qsh} and which is employed also in the present paper. The method uses the knowledge of the particle composition at LHC energies, i.e. the abundances of the different particle species within a specific interval of \pt and for a specific event class. 

To correct the distributions for secondary-particle contamination, i.e. the products of weak decays of kaons and $\Lambda$ baryons, and the particles originating from interactions in the detector material, we used the $d_{\rm xy}$ distributions of particles in data and Monte Carlo simulations.  Exploiting the differences of the $d_{\rm xy}$ distributions between primary and secondary particles, especially in the tails, the measured distributions were fitted by a linear combination of $d_{\rm xy}$ distributions (templates) for primary and secondary particles obtained from Monte Carlo simulations in different \pt bins. For INEL$\,>0$ pp collisions at $\sqrt{s}=13$\,TeV the contamination ranges from 8.5\% at $\pt=0.2$\,GeV/$c$  to 1\% for $\pt > 2$\,GeV/$c$. The contamination exhibits a small multiplicity dependence, which is below 2\%. For pp collisions at $\sqrt{s}=5.02$\,TeV, the correction factors reach similar values.

The transverse momentum spectra are also corrected for \pt resolution; the correction factor is calculated using the covariance matrix of the Kalman fit~\cite{kalman60}. The \pt (multiplicity) dependence of the correction factor is negligible~\cite{Acharya:2018qsh} (below 1\%).

Finally, the \pt spectra are corrected for the amount of signal which is missing from the yield due to the event selection (signal loss). This correction is negligible for high-multiplicity events and reaches 13\% (4\%) at $\pt=0.2$ ($\pt=1$)\,GeV/$c$ for the lowest multiplicity class based on $N_{\rm SPD\,tracklets}$.

\subsection{Spherocity studies}

The measurement of the average transverse momentum as a function of charged-particle multiplicity and spherocity is performed following a strategy close to that used in earlier publications~\cite{Aamodt:2010my,Abelev:2013bla}. The transverse momentum spectra for different multiplicity and spherocity classes are fully corrected as described in the previous section.  The average transverse momentum is then calculated from the corrected spectra as the arithmetic mean in the kinematic range $0.15<\pt<10$\,GeV/$c$ and $|\eta|<0.8$. 

To extract the correlation between \mpt and the number of primary charged particles ($N_{\rm ch}$) in $|\eta| ~ < ~ 0.8$ and for the spherocity class $S_{0}$, the following re-weighting procedure is applied to account for the experimental resolution of the measured event multiplicity ($N_{\rm m}$) and spherocity ($S_{\rm m}$):

\begin{equation}
\mpt(N_{\rm ch},S_{0})=\sum_{N_{\rm m}} \sum_{S_{\rm m}}R(N_{\rm ch},N_{\rm m})  \mpt(N_{\rm m},S_{\rm m})  R^{\prime}(S_{0},S_{\rm m}).
\label{eqso:1}
\end{equation}

This method is an extension to the one developed for the previous \mpt analysis~\cite{Aamodt:2010my}. It exploits the normalized response matrices $R$ and $R^{\prime}$ which encode  the multiplicity, and spherocity detector resolutions, respectively. The average \pt for the $S_{0}$ event class is encoded inside the inner sum, where the weights $R^{\prime}(S_{0},S_{\rm m})$ are explicitly applied to \mpt values. The resulting $\mpt(N_{\rm m},S_{0})$ is then corrected for multiplicity resolution. It is worth mentioning that the spherocity-integrated class (0\,--\,100\%) only requires the multiplicity correction. The Monte Carlo non-closure, discussed in the next section, is assigned as systematic uncertainty. 

\section{Systematic uncertainties}\label{s:4}

\subsection{Transverse momentum spectra}

The relative systematic uncertainties on the \pt spectra are summarized in Table~\ref{tab:2a}. They include the effect of the event selection based on the vertex position, which is studied by comparing the fully corrected \pt spectra obtained with alternative vertex selections: $|z|<5$\,cm and $|z|<20$\,cm. The corrections due to trigger and vertex selection were determined using the EPOS~LHC~\cite{Pierog:2013ria} event generator and the deviations with respect to the nominal values, i.e.~those obtained with PYTHIA~6, were assigned as systematic uncertainties. The same procedure was employed for the estimation of the systematic uncertainty associated to the signal loss correction. The systematic uncertainty related to the track selection was studied by varying the track cuts for which we used the variation intervals  described in Ref.~\cite{Acharya:2018qsh}. We also studied the systematic effects related to the uncertainty on the primary particle composition which is assumed for the efficiency correction. This uncertainty takes into account the extrapolation of the spectra to low \pt, the relative particle abundances at high \pt,  the uncertainties of the measured particle spectra, and the Monte Carlo assumptions on the $\Sigma^{\pm}/\Lambda$ spectra ratios. The systematic uncertainties of the correction for secondaries contamination is estimated by varying the fit model using two templates, i.e. for primaries and secondaries, or three templates, i.e.  primaries, secondaries from interactions in the detector material, and secondaries from weak decays, as well as varying the fit momentum ranges. Since we are using the same event selection and track cuts as those used in Ref.~\cite{Acharya:2018qsh},  the systematic uncertainties associated with matching efficiency, \pt resolution and material budget, are identical.

\begin{table}[h!]
\centering
  \caption{Main sources and values of the relative systematic uncertainties of  transverse momentum spectra for pp collisions at $\sqrt{s}=5.02$\,TeV. The maximum values of the uncertainties, among all the multiplicity classes, are reported for low-, intermediate-, and high-\pt intervals. Systematic uncertainties for pp collisions at $\sqrt{s}=13$\,TeV are shown inside the parentheses. The systematic uncertainty due to trigger and event selection is \pt independent and therefore it is not included in the table. It reaches $\sim 7.6$\% ($\sim 6.3$\%) for the lowest multiplicity class in pp collisions at $\sqrt{s}=5.02$\,TeV ($\sqrt{s}=13$\,TeV), and it is smaller than 0.5\% for the other multiplicity classes.}
\begin{tabular}{ p{5cm}p{2.3cm}p{2.3cm}p{2.3cm}  }
\hline
\multicolumn{4}{c}{pp collisions at $\sqrt{s}=5.02$\,TeV ($\sqrt{s}=13$\,TeV) } \\
\hline
\pt (GeV/$c$) & 0.15 & 3.0 & 10 \\
\hline
   Pileup              &  3.5\% (3.5\%)   &   4.5\% (4.5\%)   & 4.5\% (4.5\%)  \\ 
   Vertex selection              &  0.5\% (0.5\%)   &   0.5\% (0.5\%)   & 0.5\% (0.5\%)  \\
   Signal loss              &  2.3\% (1.1\%)   &   1.2\%(0.5\%)   & 0.4\% (0.7\%)  \\   
   Track selection              &  1.6\% (1.7\%)   &   3.1\% (1.7\%)   & 4.0\% (4.0\%)  \\
   Secondary particles              &  1.3\% (1.4\%)   &   1.0\% (1.0\%)   & 1.0\% (1.0\%)  \\
   Particle composition              &  2.0\% (2.0\%)   &   2.5\% (2.5\%)   & 2.0\% (2.0\%)  \\
   Tracking efficiency              &  1.0\% (1.0\%)   &   4.2\% (4.2\%)   & 4.2\% (4.2\%)  \\
   \pt resolution              &  0.0\% (0.0\%)   &   0.0\% (0.0\%)   & 0.1\% (0.1\%)  \\ 
   Material budget          &  1.5\% (1.5\%)   &   0.5\% (0.5\%)   & 0.2\% (0.2\%)  \\                   
   \hline
  Total & 5.4\% (5.1\%)     & 7.5\% (7.0\%)  & 7.7\% (7.7\%)  \\
  Total ($N_{\rm ch}$-dependent) & 4.1\% (3.5\%)     & 5.8\% (5.5\%) & 5.9\% (6.5\%)   \\
\hline
\end{tabular}
  \label{tab:2a}
\end{table}

\subsection{Average transverse momentum}

A summary of the systematic uncertainties for three multiplicity values and for different spherocity classes is shown in Table~\ref{tab:2b}. In order to estimate the systematic uncertainties of \mpt, the results of the data analysis and of the evaluation of the corrections from Monte Carlo simulations were studied considering cut variations and Monte Carlo assumptions, within reasonable limits. The effect of the track cuts on \mpt was found to be spherocity independent and of the order of 1\%. The efficiency correction is another spherocity independent contribution and it is found to be $\sim$1\%. This contribution takes into account the different particle composition in data and models, as well as the multiplicity dependence of the correction. We also studied the multiplicity dependence of the purity correction; the effect was found to be smaller than 0.5\%. The most relevant spherocity independent contribution is related to the re-weighting procedure to correct for the detector multiplicity resolution. This was quantified from the Monte Carlo non-closure, it amounts to $\sim$1.36\%,  $\sim$0.86\% and  $\sim$1.26\% for ${\rm d}N_{\rm ch}/{\rm d}\eta=1.88$, 6.25, 25.00, respectively.

The set of track cuts used to measure spherocity was also varied compared to those used for the \pt spectra analysis. The effect on the results amounted to 1\%. The most relevant contribution to the systematic uncertainties originates from the re-weighting procedure method which is used to correct for the spherocity resolution. The Monte Carlo non-closure is assigned as a systematic uncertainty. For the lowest multiplicity value, ${\rm d}N_{\rm ch}/{\rm d}\eta$=1.88, the uncertainty reaches  3.23\%, 4.55\%, and 7.06\% for the 0\,--\,10\%, 40\,--\,50\%, and 90\,--\,100\% spherocity classes, respectively. For higher multiplicities, e.g. ${\rm d}N_{\rm ch}/{\rm d}\eta$=25.0, the Monte Carlo non-closure amounts to 0.57\%, 1.07\%, and 2.01\%  for the 0\,--\,10\%, 40\,--\,50\%, and 90\,--\,100\% spherocity classes, respectively. As expected from the detector response, the most relevant effects are observed for low-multiplicity events in particular for the isotropic classes. As will be seen later, in Monte Carlo jet-like events, the average \pt shows a strong change with multiplicity at ${\rm d}N_{\rm ch}/{\rm d}\eta\sim7$. This effect increases the size of the uncertainty (Monte Carlo non-closure) in that multiplicity interval. This is the dominant contribution to the systematic uncertainties and covers the largest variations observed between data and PYTHIA~8 (version 8.212)~\cite{Sjostrand:2014zea} (tune Monash 2013~\cite{Skands:2014pea}).

The model dependence is also checked by using events simulated with PYTHIA~8 and EPOS~LHC which include the particle transport through the detector. The corrections were calculated using these simulations and the maximum variation with respect to the nominal values (using PYTHIA~6 simulations) are below 1\%.

\begin{table}[h!]
\centering
  \caption{Main sources and values of the relative systematic uncertainties on the average \pt for different spherocity classes. The three quoted values (for each contribution) correspond to ${\rm d}N_{\rm ch}/{\rm d}\eta=1.88$, 6.25,  and 25.0, respectively.}
\begin{tabular}{ p{3.8cm}p{2.4cm}p{2.4cm}p{2.4cm}p{2.4cm}| }
\hline
\multicolumn{5}{c}{Spherocity-dependent contributions} \\
\hline
Spherocity class                            & 0\,--\,100\%                & 0\,--\,10\%                                 &  40\,--\,50\%                             & 90\,--\,100\%                             \\
\hline   
Model dep.    (\%)                          & 0.5, 0.7, 0.2     & 0.4, 0.4, 0.6                   & 0.6, 0.5, 0.3                  & 0.9, 0.8, 0.2      \\
Sec. particles (\%) &0.2, 0.3, 1.2 & 1.0, 1.5, 1.9 & 0.3, 0.3, 1.1 & 0.6, 0.3, 0.9 \\ 
Ev. selection (\%)  &2.2, 0.0, 0.0 & 1.9, 1.4, 0.4 & 1.3, 0.4, 0.0 & 1.3, 0.1, 0.00 \\ 
$S_{\rm o}$ res. corr.     (\%)         &        \multicolumn{1}{c}{\multirow{2}{*}{na}}        & 3.2, 5.4, 0.6                   & 4.6, 2.4, 1.1                  & 7.1, 3.7, 2.0        \\
$S_{\rm o}$ track cuts  (\%)        &   \multicolumn{1}{c}{}     &  \multicolumn{3}{c|}{1.00} \\ 
\hline
\multicolumn{5}{c}{Spherocity-independent contributions} \\
\hline
$N_{\rm ch}$ res. corr (\%) & \multicolumn{4}{c|}{1.4, 0.9, 1.3}       \\
\pt track cuts (\%)  &  \multicolumn{4}{c|}{0.8, 0.9, 1.2}   \\
Efficiency corr.  (\%) &   \multicolumn{4}{c|}{0.4, 0.2, 0.2}   \\
Particle composition (\%) &  \multicolumn{4}{c|}{1.0, 1.0, 1.0}   \\
$N_{\rm ch}$ dep. eff. corr. (\%) &  \multicolumn{4}{c|}{0.5, 0.7, 0.9}    \\
$N_{\rm ch}$ dep. sec. corr.  (\%) & \multicolumn{4}{c|}{0.2, 0.1, 0.1}   \\
\hline
$S_{\rm o}$-dep. total (\%) &2.2, 0.8, 1.2 & 4.0, 5.8, 2.3 & 4.9, 2.7, 1.9 & 7.3, 3.9, 2.4 \\
$S_{\rm o}$-indep. total (\%)    & \multicolumn{4}{c|}{2.0, 1.8, 2.2}   \\
Total  (\%)         &3.0, 2.0, 2.5 & 4.5, 6.1, 3.2 & 5.3, 3.2, 3.0 & 7.6, 4.3, 3.3 \\
\hline
 \end{tabular}
  \label{tab:2b}
\end{table}

\begin{figure}[htb]
\centerline{
\includegraphics[width=0.7\columnwidth]{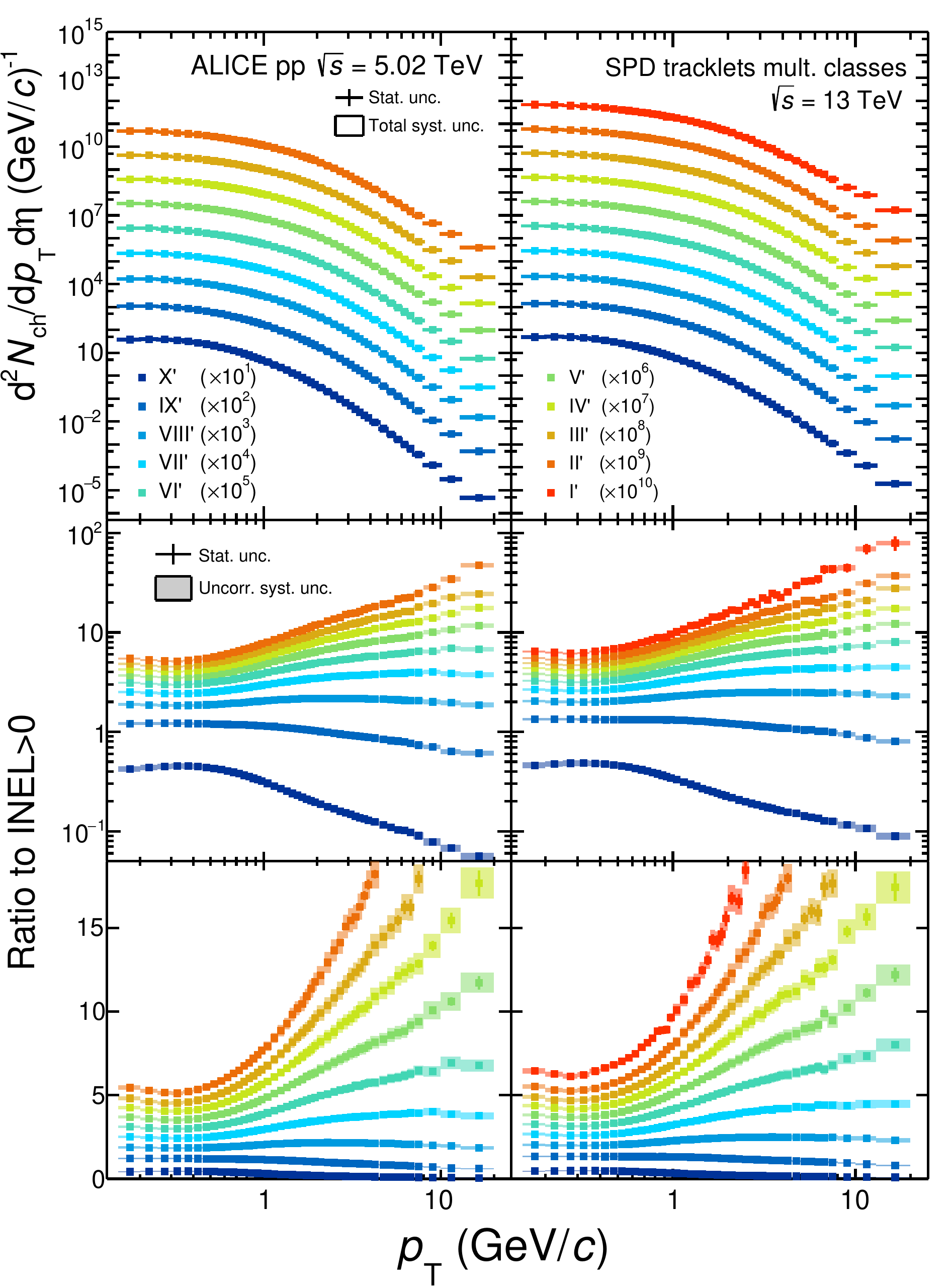}}
\caption{Transverse momentum distributions of charged particles for multiplicity classes selected using SPD tracklets in $|\eta|<0.8$. Results for pp collisions at $\sqrt{s}=5.02$ and 13\,TeV are shown in the left and right panels, respectively. Statistical and total systematic uncertainties are shown as error bars and boxes around the data points, respectively. In the middle panels, ratios of multiplicity dependent spectra to INEL$\,>0$ are shown in logarithmic scale. In the bottom panels we show the ratios in a linear scale to illustrate the dramatic behavior of the ratios. The systematic uncertainties on the ratios are obtained by considering only contributions uncorrelated across multiplicity. The spectra are scaled to improve the visibility.}
\label{fig:1}
\end{figure}

\begin{figure}[htb]
\centerline{
\includegraphics[width=0.7\columnwidth]{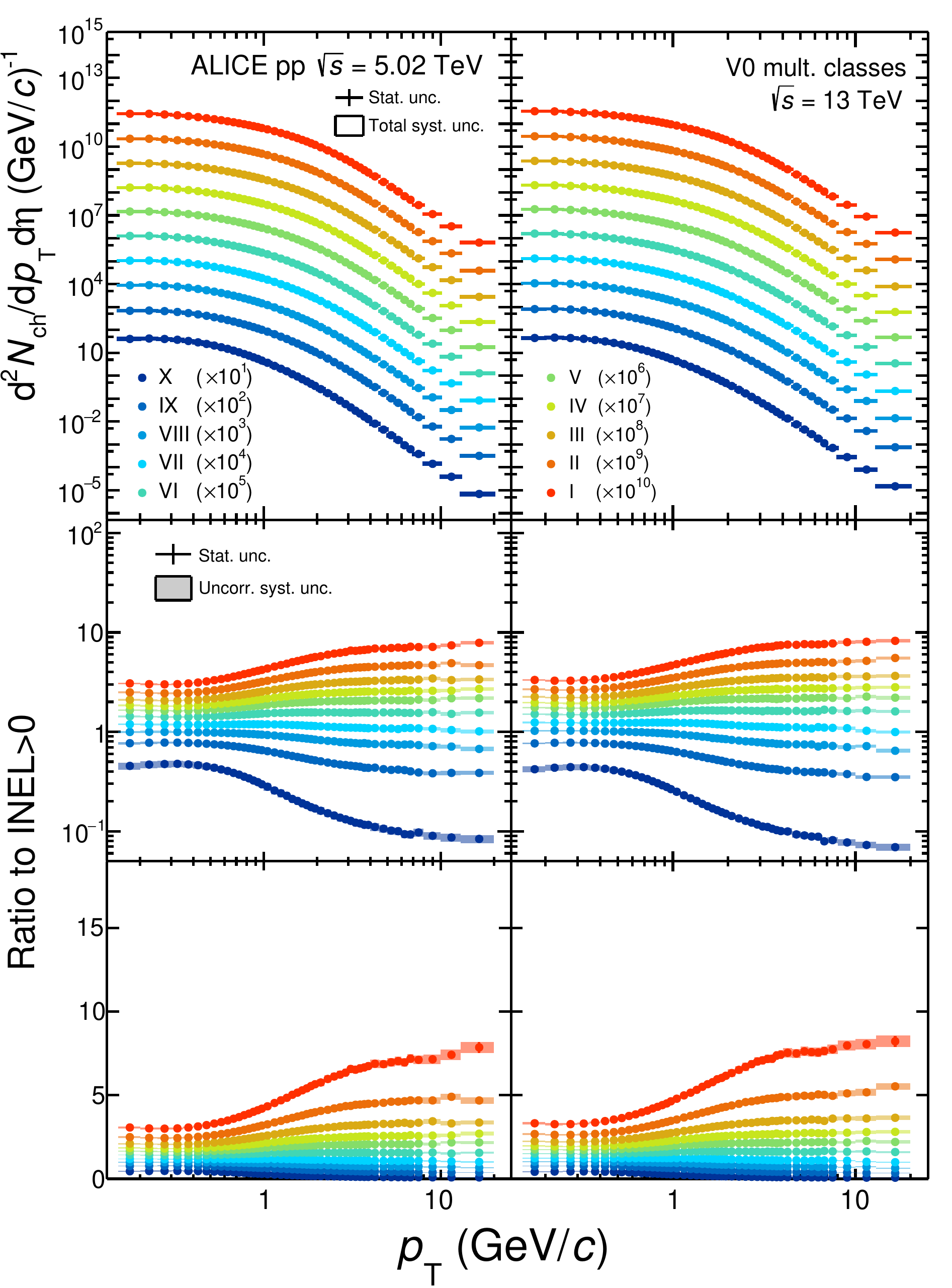}}
\caption{Transverse momentum distributions of charged particles for different V0M multiplicity classes. Results for pp collisions at $\sqrt{s}=5.02$ and 13\,TeV are shown in the left and right panels, respectively. Statistical and total systematic uncertainties are shown as error bars and boxes around the data points, respectively. In the middle panels, ratios of multiplicity dependent spectra to INEL$\,>0$ are shown in logarithmic scale. The systematic uncertainties on the ratios are obtained by considering only contributions uncorrelated across multiplicity. The spectra are scaled to improve the visibility.}
\label{fig:2}
\end{figure}

\begin{figure}[htb]
\centerline{
\includegraphics[width=0.5\columnwidth]{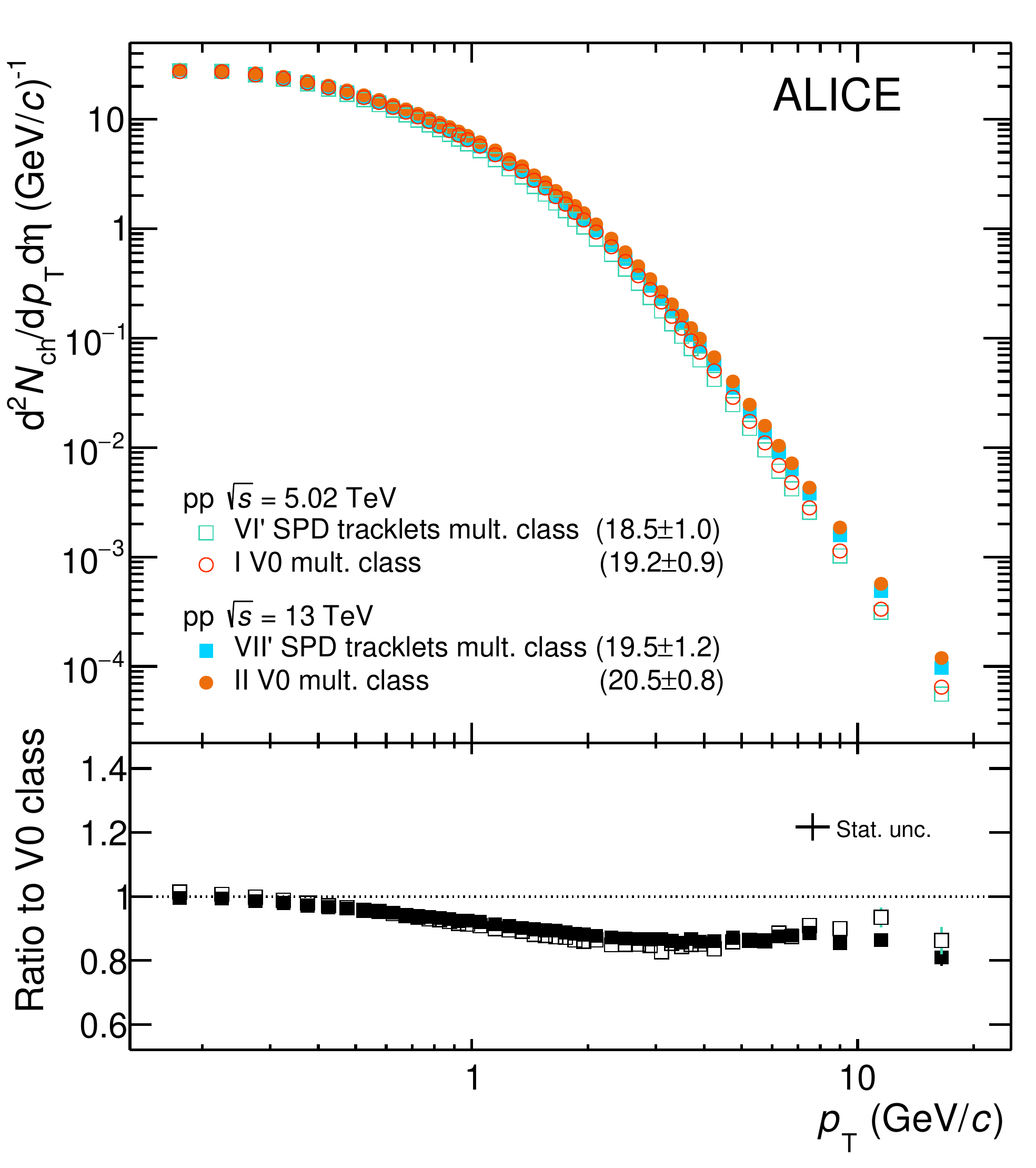}}
\caption{Transverse momentum distributions of charged particles for high-multiplicity ($\langle {\rm d}N_{\rm ch}/{\rm d}\eta \rangle \approx20$) pp collisions at $\sqrt{s}=5.02$ (empty markers) and 13\,TeV (full markers). Results for V0-based (squares) and SPD-based (circles) multiplicity estimators are shown. The bottom panel shows the \pt spectrum obtained using the V0-based multiplicity estimator normalized to that using the SPD-based multiplicity estimator. Only statistical uncertainties are shown as error bars.}
\label{fig:new}
\end{figure}

\begin{figure}[htb]
\centerline{
\includegraphics[width=0.99\columnwidth]{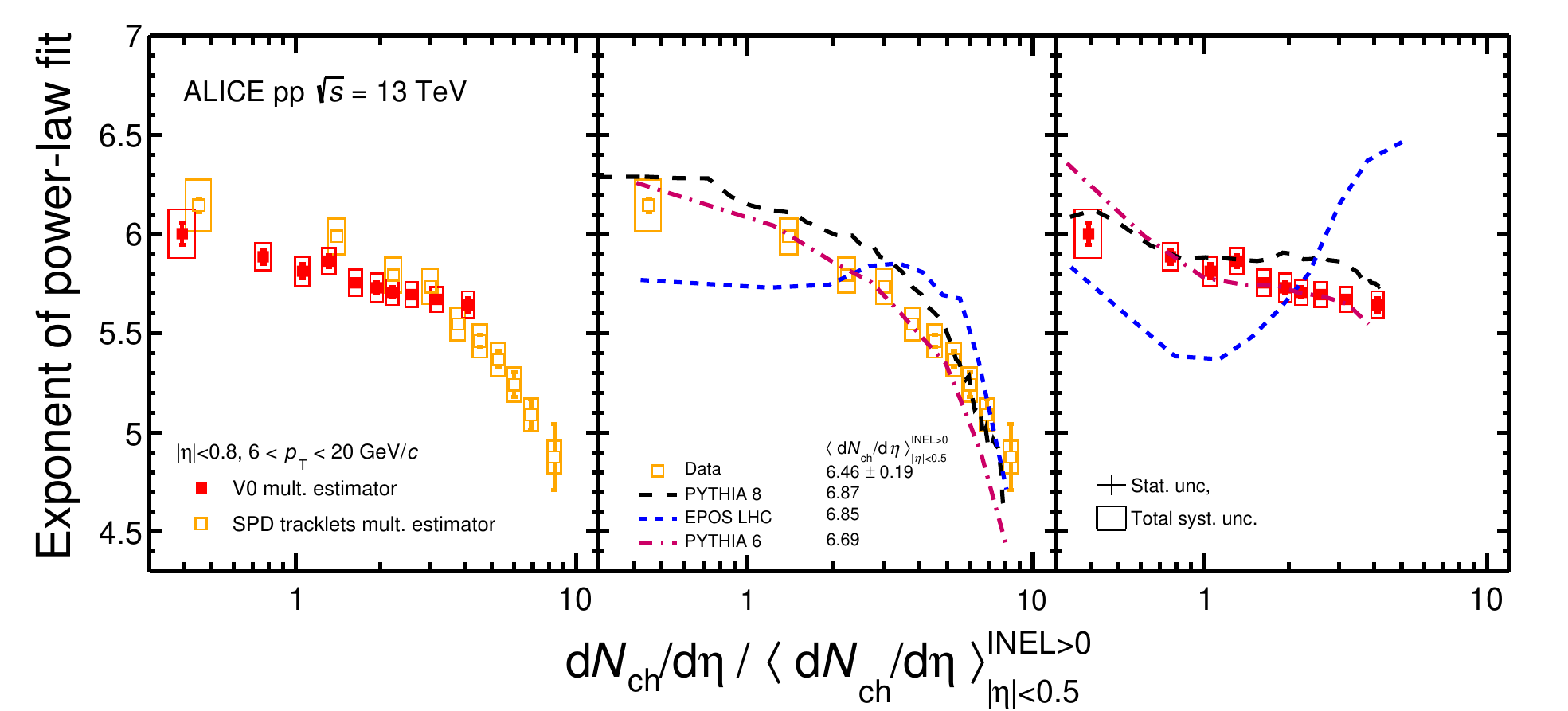}}
\caption{Evolution of the spectral shape of the transverse momentum distribution as a function of charged-particle multiplicity. The spectral shape is characterized by the exponent of the power-law function which fits the high-\pt part ($\pt>6$\,GeV/$c$) of the invariant yields. Results for pp collisions at $\sqrt{s}=13$\,TeV are shown. A comparison of the two multiplicity estimators discussed in this paper is shown in the left panel. Comparisons with Monte Carlo generators predictions are shown in the middle and right panels. Statistical and total systematic uncertainties are shown as error bars and boxes around the data points, respectively.}
\label{fig:3}
\end{figure}

\section{Results}\label{s:5}

\subsection{Transverse momentum spectra as a function of charged-particle multiplicity}

The \pt distributions of charged particles, measured in $|\eta|<0.8$ for pp collisions at $\sqrt{s}=5.02$ and 13\,TeV, are shown in Fig.~\ref{fig:1} for the different multiplicity classes selected using the estimator based on $N_{\rm SPD\,tracklets}$. The bottom panels depict the ratios to the \pt distribution of the INEL$\,>0$ event class. The features of the spectra, i.e. the change of the spectral shape going from low- to high-multiplicity values, are qualitatively the same for both energies. The only significant difference is the multiplicity reach which is higher at 13\,TeV than that at 5.02\,TeV. In the following we discuss the results for pp collisions at the highest energy.  As shown in Fig.~\ref{fig:1}, the \pt spectra become harder as the multiplicity increases, which contributes to the increase of the average transverse momentum with  multiplicity.  The ratios to the INEL$>0$ \pt distribution exhibit two distinct behavior. While at low \pt ($<0.5$\,GeV/$c$) the ratios exhibit a modest \pt dependence, for $\pt>0.5$\,GeV/$c$ they strongly depend on multiplicity and \pt.

\begin{figure}[htb]
\centerline{
\includegraphics[width=0.5\columnwidth]{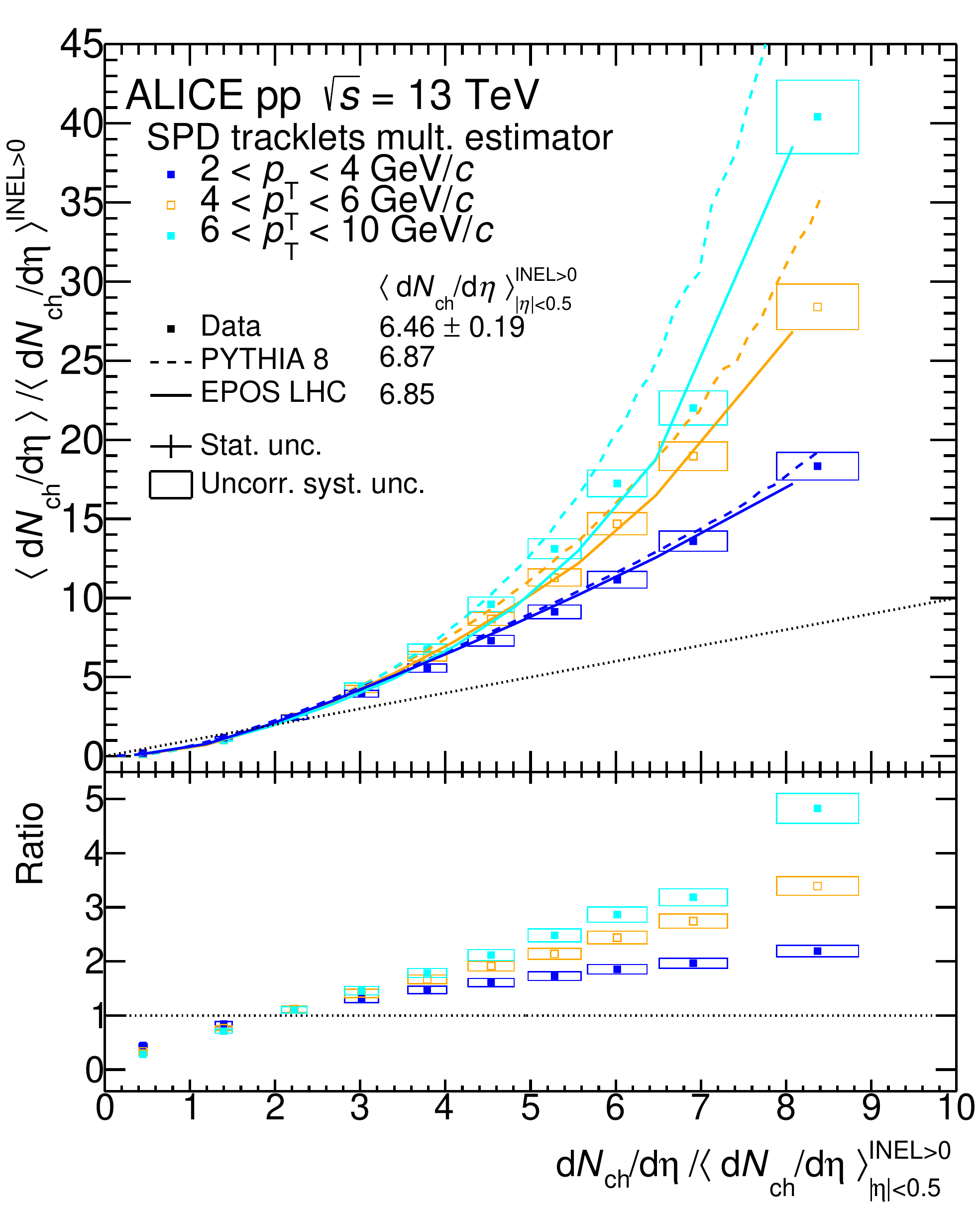}}
\caption{Self-normalized yields of charged particles integrated over different \pt intervals: $2<\pt<4$\,GeV/$c$, $4<\pt<6$\,GeV/$c$, and $6<\pt<10$\,GeV/$c$. The integrated yields for pp collisions at $\sqrt{s}=13$\,TeV are shown as a function of charged-particle density at mid-pseudorapidity. Statistical and uncorrelated (across multiplicity) systematic uncertainties are shown as error bars and boxes around the data points, respectively. Statistical uncertainties are negligible compared to systematic uncertainties. Data are compared with PYTHIA~8 (dashed lines) and EPOS~LHC (solid lines).  The dotted line is drawn to see the differences between data and the linear dependence. Deviations of data from the linear trend are shown in the bottom panel. }
\label{fig:4}
\end{figure}

Figure~\ref{fig:2} shows the multiplicity dependent \pt spectra using a multiplicity selection based on the V0M amplitude. Results for pp collisions at $\sqrt{s}=5.02$ and 13\,TeV are shown. The average multiplicity values are significantly smaller than those reached with the mid-pseudorapidity estimator (based on $N_{\rm SPD\,tracklets}$). For  example, in pp collisions at $\sqrt{s}=13$\,TeV, while the average charged-particle multiplicity density amounts to 56.55 for the highest $N_{\rm SPD\,tracklets}$ class, it only reaches 27.61 for the highest V0M multiplicity class. We note that for similar average particle densities, e.g.~the multiplicity classes II (V0M) and VII' (SPD tracklets)  in pp collisions at $\sqrt{s}=13$\,TeV, the ratios measured using the V0M amplitude and the $N_{\rm SPD\,tracklets}$ are similar. The comparison of the \pt spectra for these multiplicity classes is shown in Fig.~\ref{fig:new}.  We observe that for transverse momentum below 0.5\,GeV/$c$, the spectra exhibit the same shape. For transverse momenta within 0.5--3\,GeV/$c$ the spectra for the multiplicity class II is harder than that for the VII'' class. At higher \pt, the spectral shapes are the same, but the yield of the class II is $\sim$15\% higher than that for the VII' class. Similar results are obtained if we compare the multiplicity classes I and VI' for pp collisions at 5.02\,TeV.
 
Commonly, the particle production is characterized by quantities like integrated yields, or any fit parameter of the curve extracted from fits to the data, for example, the so-called inverse slope parameter reported by ALICE in Ref.~\cite{Abelev:2013vea}.  This facilitates the visualization of the evolution of the particle production as a function of multiplicity and the comparison among different colliding systems. Several publications have adopted this strategy for soft  ($\pt<2$\,GeV/$c$)~\cite{Abelev:2013haa,ALICE:2017jyt,Abelev:2013bla} physics and others to describe the particle production for intermediate and high \pt ($2 \leq \pt<20$\,GeV/$c$)~\cite{Abelev:2012cn}. It is interesting and important to define a common quantity to compare the shape of the high-\pt part of the spectra of different particle species and collision systems. The natural choice is fitting a power-law function ($\alpha \times \pt^{-n}$) to the invariant yield and studying the multiplicity dependence of the exponent ($n$) extracted from the fit. Figure~\ref{fig:3} illustrates the results considering particles with transverse momentum within 6--20\,GeV/$c$ for pp at $\sqrt{s}=13$\,TeV. It is worth mentioning that within uncertainties the power-law function describes rather well the data in that \pt interval. Similarly, the \pt spectra simulated with the different generators are well described (within 2\%) by the power-law function.

Within uncertainties, going from low to high multiplicity $n$ decreases taking values from 6 to 5, respectively. A similar behavior has been reported for heavy-ion collisions~\cite{Mishra:2018pio}. Moreover,  the results using the two multiplicity estimators are consistent within the overlapping multiplicity interval. This result is consistent with that shown in Fig.~\ref{fig:new}. PYTHIA~6 and 8 simulations describe the trends very well, but a strong deviation between EPOS~LHC and data is observed. In PYTHIA~8, it has been shown that the number of high-\pt jets increases with event multiplicity. Moreover, for a given event multiplicity and fixed jet \pt, the high-\pt tails of the charged-particle spectra are very similar in low- and high-multiplicity events~\cite{Ortiz:2016kpz}. Therefore, based on PYTHIA~8 studies, the reduction of the power-law exponent with increasing multiplicity can be attributed to an increasing number of high-\pt jets.   

As pointed out above, the ratios to the INEL$\,>0$ \pt distributions for $\langle {\rm d}N_{\rm ch}/{\rm d}\eta \rangle\lesssim25$ exhibit a weak \pt-dependence for $\pt>4$\,GeV/$c$. This applies to both energies and to all multiplicity estimators.  To illustrate better the behaviour of the yields at high momenta,  we adopted a representation previously used for heavy-flavour hadrons~\cite{Adam:2015ota} to point out to the similarities between the two results. The trend at high-\pt is highlighted in Fig.~\ref{fig:4}, which shows the integrated yields for three transverse momentum intervals ($2<\pt<10$\,GeV/$c$, $4<\pt<10$\,GeV/$c$, and $6<\pt<10$\,GeV/$c$) as  a  function  of  the average mid-pseudorapidity  multiplicity.   Both the  charged-particle yields and  the average multiplicity are self-normalized, i.e.~they are divided by their average value for the INEL$\,>0$ sample.  The high-\pt ($>4$\,GeV/$c$) yields of charged particles increase faster than the charged-particle multiplicity, while the increase is smaller when we consider lower-\pt particles. The trend of the data is qualitatively well reproduced by PYTHIA~8, but for $\pt>6$\,GeV/$c$ the model significantly overestimates the ratio by a factor larger than 1.5. Although the shapes of the spectra (characterized by $n$) are not well reproduced  by EPOS~LHC, the model gives the best description of the self-normalized yields.  Despite the large uncertainties, it is clear the data show a non-linear increase.

\newpage

\subsection{ Double-differential study of the average transverse momentum }

\begin{figure}[htb]
\centerline{
\includegraphics[width=0.5\columnwidth]{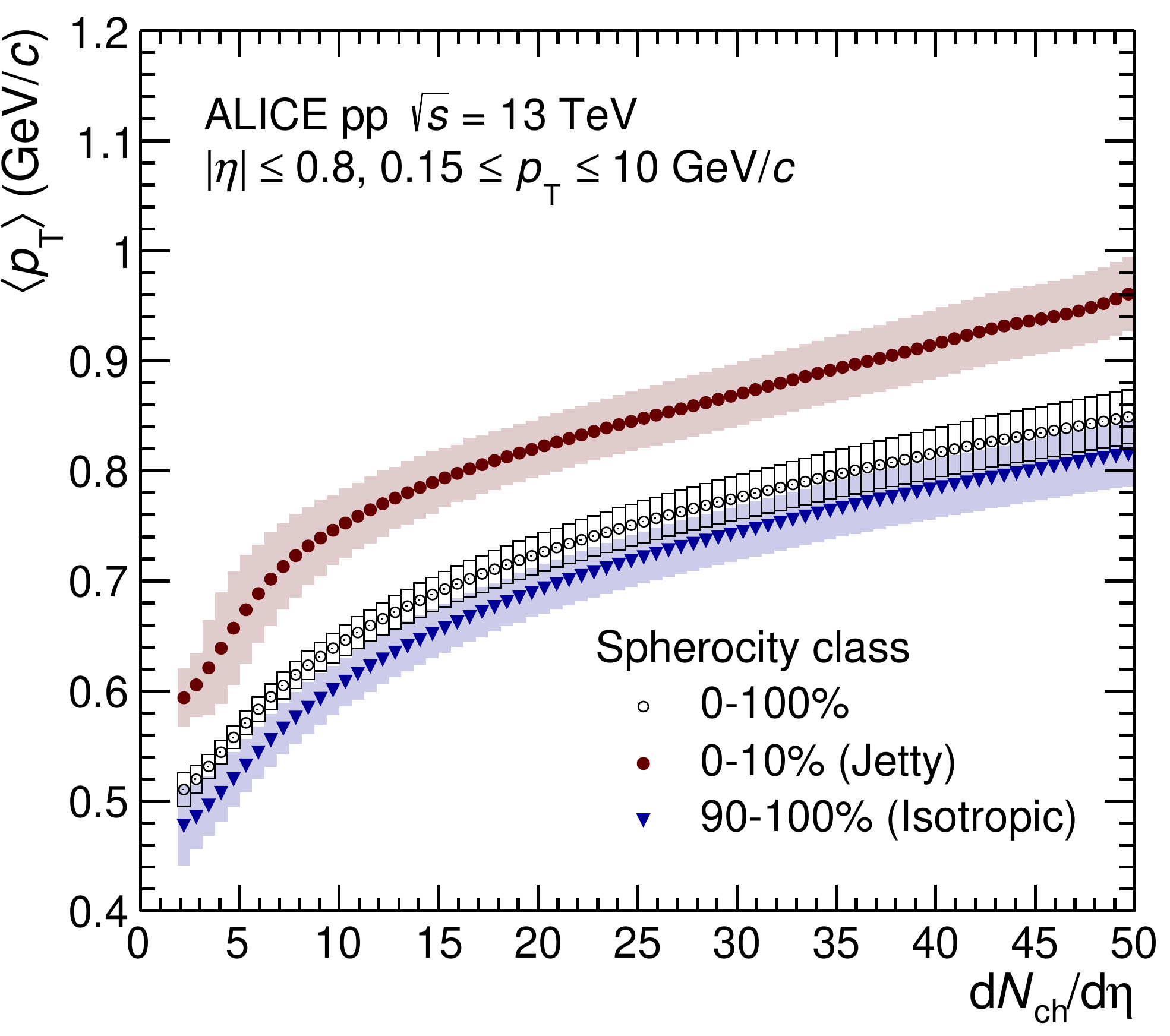}}
\caption{Average transverse momentum as a function of event multiplicity in pp collisions at $\sqrt{s}=13$\,TeV. Results for the spherocity-integrated case (0\,--\,100\%) are contrasted with the measurements for the most jet-like (0\,--\,10\%) and isotropic (90\,--\,100\%) events. Statistical uncertainties (error bars) are negligible compared to  systematic uncertainties (boxes around the data points).}
\label{so:3}
\end{figure}

The spherocity-integrated average \pt as a function of ${\rm d}N_{\rm ch}/{\rm d}\eta$ for pp collisions at $\sqrt{s}=13$\,TeV is shown in Fig.~\ref{so:3}. In accordance with measurements at lower energies~\cite{Abelev:2013bla}, the \mpt increases with ${\rm d}N_{\rm ch}/{\rm d}\eta$. In PYTHIA~8 the effect is enhanced by color reconnection, which allows the interaction among partons originating from multiple semi-hard scatterings via color strings. The minimum-bias data are compared with analogous measurements for the most jet-like structure (0\,--\,10\%) and isotropic (90\,--\,100\%) event classes. Studying observables as a function of spherocity reveals interesting features. On one hand, for isotropic events the average $p_{\rm T}$ stays systematically below the spherocity-integrated \mpt over the full multiplicity range; on the other hand, for jet-like events the \mpt is higher than that for spherocity-integrated events.  Moreover, within uncertainties the overall shape of the correlation, i.e. a steep linear rise below  ${\rm d}N_{\rm ch}/{\rm d}\eta=10$ followed by a less steep but still linear rise above, is not spherocity-dependent.

\begin{figure}[htb]
\centerline{
\includegraphics[width=1.0\columnwidth]{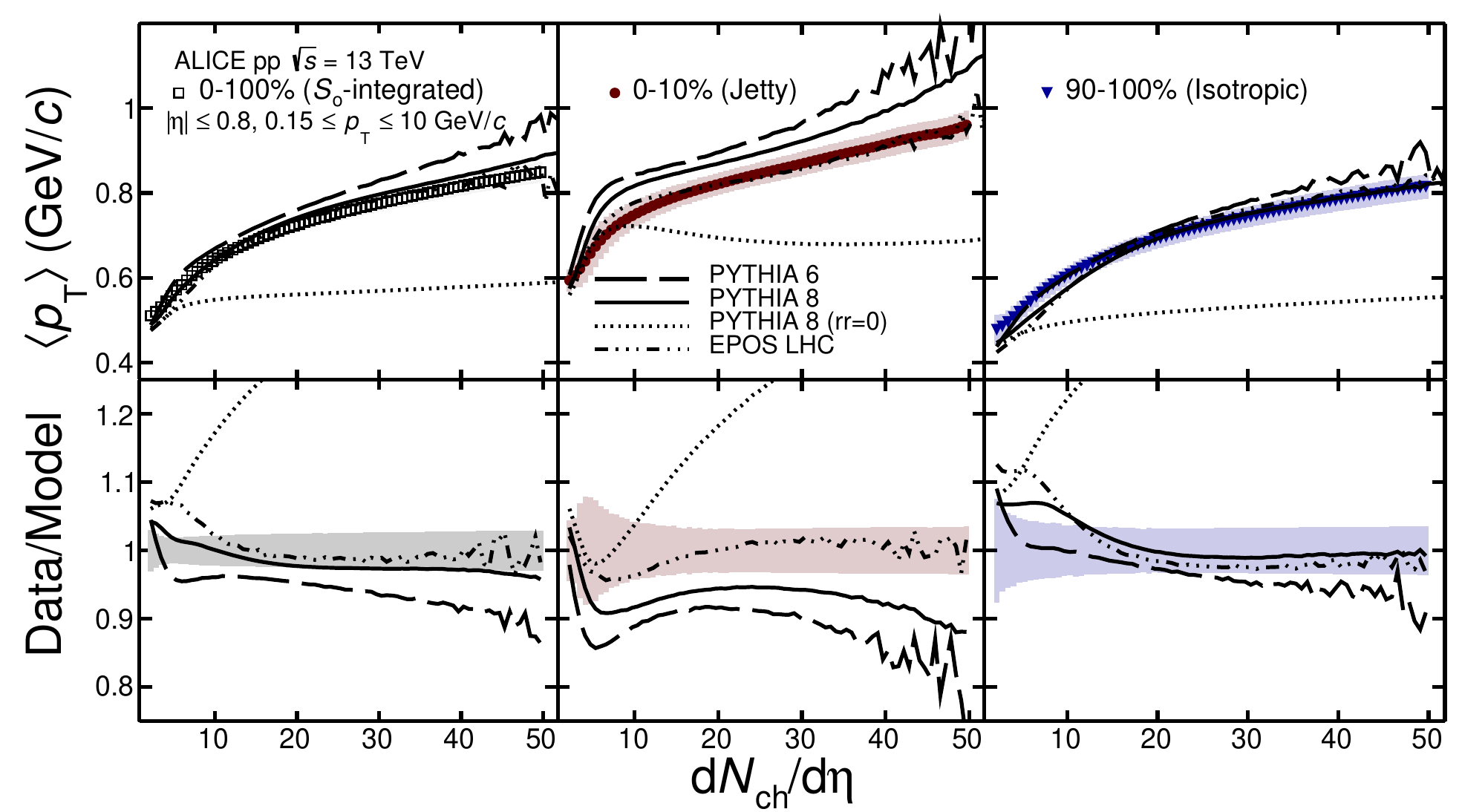}}
\caption{Average transverse momentum as a function of event multiplicity in pp collisions at $\sqrt{s}=13$\,TeV. Results for the spherocity-integrated case (0\,--\,100\%), the most jet-like (0\,--\,10\%) and isotropic (90\,--\,100\%) events are compared with Monte Carlo predictions. Predictions of PYTHIA~8 with and without (null reconnection range, $\rm{rr}=0$) color reconnection, as well as PYTHIA~6 and EPOS~LHC are displayed. Statistical uncertainties (error bars) are negligible compared to  systematic uncertainties (shaded area around the data points). Data to model ratios are shown in the bottom panel. The color band around unity represents the systematic uncertainty.}
\label{so:4}
\end{figure}

Figure~\ref{so:4} shows that within uncertainties, PYTHIA~8 with color reconnection gives an adequate description of the spherocity-integrated event class. It is worth mentioning that color reconnection was originally introduced to explain the rise of \mpt with multiplicity~\cite{Sjostrand:1987su}. However, PYTHIA~6 shows a steeper rise of \mpt with ${\rm d}N_{\rm ch}/{\rm d}\eta$ than that seen in data. The Perugia~2011 tune relies on Tevatron and SPS minimum-bias data, while the Monash tune was constrained using the early LHC measurements~\cite{Skands:2014pea}. The comparison of data with EPOS~LHC is also shown. Clearly, the quantitative agreement is as good as that achieved by PYTHIA~8. The EPOS~LHC model uses a different approach in order to simulate the hadronic interactions. Namely, the model considers a collective hadronization which depends only on the geometry and the density~\cite{Pierog:2013ria}.  

For the 0\,--\,10\% and 90\,--\,100\% spherocity classes, Fig.~\ref{so:4} also shows comparisons between data and Monte Carlo generators (PYTHIA~6, PYTHIA~8 and EPOS~LHC). It is worth mentioning that we also used spherocity percentiles in all the Monte Carlo event generators reported in this paper because their spherocity distributions do not differ much from those measured in data. For further Monte Carlo comparisons the spherocity binning which was used in the analysis is provided as HEP data.  In low-multiplicity events (${\rm d}N_{\rm ch}/{\rm d}\eta<10$), the deviations between data and PYTHIA~8 (without color reconnection) are smaller and larger respectively for the 0\,--\,10\% and 90\,--\,100\% spherocity classes than those seen for the 0\,--\,100\% spherocity class. The effect could be a consequence of the reduction of color reconnection contribution in events containing jets surrounded by a small underlying event activity. For isotropic events the three models quantitatively describe the correlation. Even for PYTHIA~6, the size of the discrepancy which was pointed out for the spherocity-integrated event class is reduced. On the contrary, for jet-like events both PYTHIA~6 and 8 exhibit a larger disagreement with the data. These models produce three distinct multiplicity regions, for ${\rm d}N_{\rm ch}/{\rm d}\eta\lesssim7$ the models give a steeper rise of \mpt than data. Within the intermediate multiplicity interval ($7\lesssim{\rm d}N_{\rm ch}/{\rm d}\eta\lesssim25$), the slope of \mpt given by models is more compatible with that seen in data, although the models overestimate the average \pt. While in data the average \pt increases at a constant rate with multiplicity for ${\rm d}N_{\rm ch}/{\rm d}\eta \gtrsim 7$, PYTHIA~6 and 8 shows a third change of the slope of \mpt, observed for ${\rm d}N_{\rm ch}/{\rm d}\eta \gtrsim 25$.  The data to model ratio indicates a discrepancy larger than 10\%, which is larger than the systematic uncertainties associated to \mpt in that multiplicity interval.

\begin{figure}[htb]
\centerline{
\includegraphics[width=0.5\columnwidth]{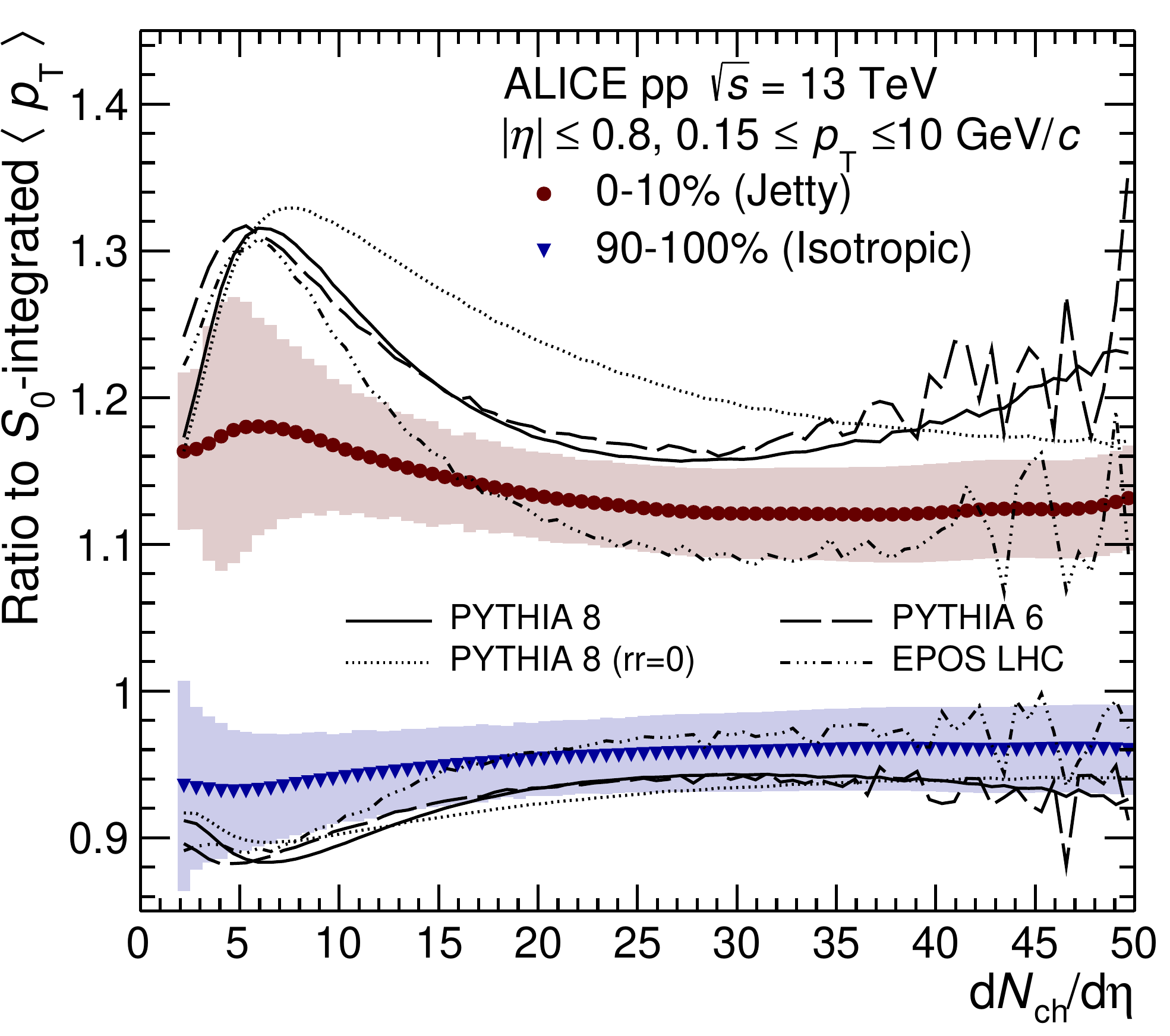}}
\caption{Average \pt of jet-like (circles) and isotropic (triangles) events normalized to that for the spherocity-integrated event class. The measurements are compared with different Monte Carlo generators. Statistical uncertainties (error bars) are negligible compared to  systematic uncertainties (boxes around the data points).}
\label{so:5}
\end{figure}

In order to study the details of the changes of the functional form of $\mpt(N_{\rm ch})$ due to the spherocity selection, Fig.~\ref{so:5} shows the average \pt of jet-like and isotropic events normalized to that for the spherocity-integrated event class. For jet-like events, the data exhibit a hint of a modest peak at ${\rm d}N_{\rm ch}/d\eta\sim7$, which is not significant if we consider the size of the systematic uncertainties. Moreover, within uncertainties the ratio remains constant for ${\rm d}N_{\rm ch}/{\rm d}\eta \gtrsim 25$. EPOS~LHC describes rather well the high-multiplicity behavior, however, it overestimates the peak. PYTHIA~6 and 8 show the worst agreement with the data. In this representation, the three distinct regions, which were described before are highlighted. In PYTHIA~8, the peak (at  ${\rm d}N_{\rm ch}/d\eta\sim7$) in jet-like events is caused by particles with transverse momentum above 2\,GeV/$c$. The size of the peak is determined by particles with $\pt>5-6$\,GeV/$c$. In contrast, data do not show a significant peak structure for any specific transverse momentum interval. We also varied the upper \pt ($0.15 < \pt <  \pt^{\rm max}$) limit ($\pt^{\rm max} = 10$ GeV/$c$ is the default) and studied the effect  on the extracted $\langle p_{\rm T} \rangle$. The $\langle p_{\rm T}\rangle$ remains constant within uncertainties for $4 < \pt^{\rm max} < 10$\,GeV/$c$ in data and for $6 < \pt^{\rm max} < 10$\,GeV/$c$ in PYTHIA~8. For $\pt^{\rm max}=2$\,GeV/c the $\langle p_{\rm T} \rangle$ decreases by 23\% (29\%) in data (PYTHIA~8) compared to $ \pt^{\rm max}=10$\,GeV/$c$. The relative difference of  $\langle p_{\rm T} \rangle$ between data and PYTHIA~8 amounts to 9\% (4\%) for $\pt^{\rm max}=2$\,GeV/$c$ ($\pt^{\rm max}=10$\,GeV/$c$). The results suggest that the power-law tail produces a smaller impact on data than in PYTHIA~8. A similar ratio for isotropic events shows a smaller structure at ${\rm d}N_{\rm ch}/d\eta\sim7$. This effect is well reproduced by all models.

\begin{figure}[htb]
\centerline{
\includegraphics[width=1.0\columnwidth]{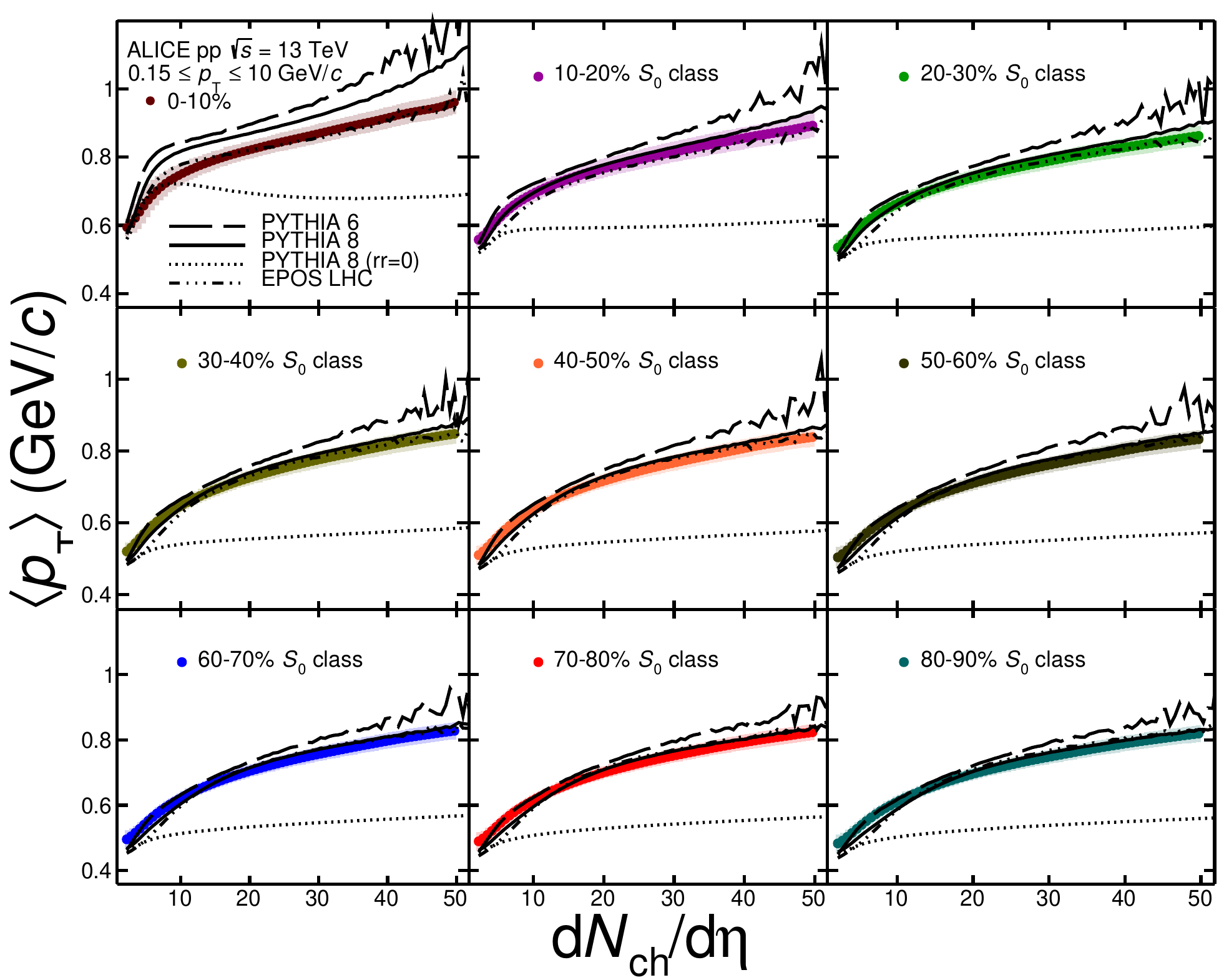}}
\caption{Average transverse momentum as a function of event multiplicity in pp collisions at $\sqrt{s}=13$\,TeV. Results for nine spherocity classes are compared with Monte Carlo predictions. Statistical uncertainties (error bars) are negligible compared to  systematic uncertainties (shaded area around the data points).}
\label{so:6}
\end{figure}

\begin{figure}[htb]
\centerline{
\includegraphics[width=1.0\columnwidth]{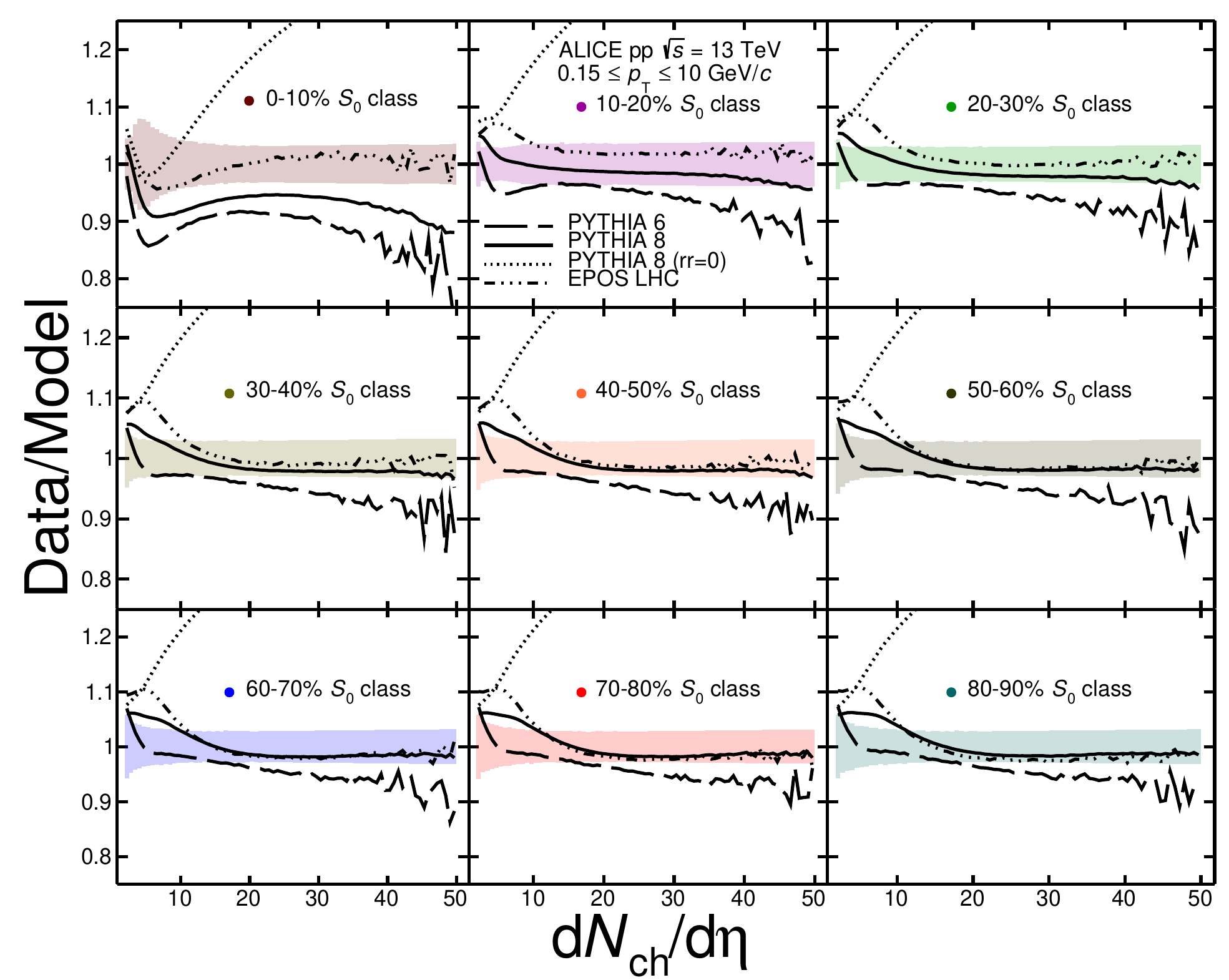}}
\caption{Data to model ratios as a function of event multiplicity in pp collisions at $\sqrt{s}=13$\,TeV. Results for nine spherocity classes are shown. Systematic uncertainties are displayed as shaded areas around unity.}
\label{so:7}
\end{figure}

Finally, we also examined the evolution of $\mpt(N_{\rm ch})$ going from the most jet-like to the most isotropic event classes. Figure~\ref{so:6} shows the spherocity-dependent $\mpt(N_{\rm ch})$ in data and models, the data to model ratios are displayed in Fig.~\ref{so:7}. The difference between the 0\,--\,10\% and 10\,--\,20\% spherocity classes is smaller for data and EPOS~LHC than for PYTHIA~6 and 8. Moreover, within uncertainties PYTHIA~8 describes rather well the data for the 10\,--\,20\% spherocity class. This contrasts with the disagreement between the model and data for the 0\,--\,10\% spherocity class. Other features in PYTHIA~6 and 8 are the reduction of the bump at ${\rm d}N_{\rm ch}/d\eta\sim7$ and the disappearance of a third rise of the \mpt for ${\rm d}N_{\rm ch}/d\eta \gtrsim 25$ when one goes from the 0\,--\,10\% to the 10\,--\,20\% spherocity classes. The agreement among models and data for the 20\,--\,100\% spherocity classes is similar to that observed for the 10\,--\,20\% spherocity class. Within uncertainties, PYTHIA~8 and EPOS~LHC qualitatively describe the data for ${\rm d}N_{\rm ch}/{\rm d}\eta \gtrsim 10$, while PYTHIA~6 overestimates the average \pt.

From previous LHC studies we know that the production cross section of jets in high-multiplicity pp collisions is smaller in data than predicted from the Monte Carlo generators~\cite{Abelev:2012sk,Chatrchyan:2013ala,Aad:2012fza}. Therefore, a possible interpretation is that the low-momentum partons, color connected with higher momentum ones (jets), would produce an overall increase of the hadron transverse momentum. This would affect more the low-\pt part of the spectrum associated with jet-enriched samples, which are achieved by requiring low-spherocity values. The incorporation of these new observables into the PYTHIA~8 tuning could be a challenge because, on one hand, the color reconnection has to be reduced to describe the low-$S_{0}$ data; on the other hand, the variation should not be too large because the good description of the spherocity-integrated and isotropic classes could be affected.

\section{Summary and conclusions}\label{s:6}

In this paper, we have reported the transverse momentum spectra of inclusive charged particles in pp collisions at $\sqrt{s}=5.02$ and 13\,TeV. The measurements were performed in the kinematic range of  $|\eta|<0.8$ and $\pt>0.15$\,GeV/$c$. The particle production was studied as a function of event multiplicity quantified by two estimators, one based on the number of SPD tracklets within $|\eta|<0.8$, and the second one based on the multiplicity in the V0 forward detector (V0M amplitude). For similar average charged-particle densities, the particle production above $\pt=1$\,GeV/$c$ is higher in pp collisions at $\sqrt{s}=13$\,TeV than at $\sqrt{s}=5.02$\,TeV. For a fixed center-of-mass energy,  particle production above $\pt=0.5$\,GeV/$c$ exhibits a remarkable multiplicity dependence. Namely, for transverse momenta below 0.5\,GeV/$c$, the ratio of the multiplicity dependent spectra to those for INEL$\,>0$ pp collisions is rather constant, and for higher momenta, it shows a significant \pt dependence. The behavior observed for each of the two multiplicity estimators are consistent within the $\langle {\rm d}N_{\rm ch}/{\rm d}\eta \rangle$ interval defined by the V0M multiplicity estimator, which gives a $\langle {\rm d}N_{\rm ch}/{\rm d}\eta \rangle$ reach of $\sim$25. For the highest V0M multiplicity class, the ratio increases going from $\pt=0.5$\,GeV/$c$ up to $\pt\approx4$\,GeV/$c$, then for higher \pt, it shows a smaller increase.

The particle production at high transverse momenta is characterized by the exponent of a power-law function which is fitted to the invariant yield considering particles with $6<\pt<20$\,GeV/$c$. Within that \pt interval, the power-law function describes rather well the \pt spectra. In concordance to the ratios discussed above, within uncertainties, the functional form of  $n$ as a function of $\langle {\rm d}N_{\rm ch}/{\rm d}\eta \rangle$ is the same for the two multiplicity estimators used in this analysis. Moreover, $n$ is found to decrease with $\langle {\rm d}N_{\rm ch}/{\rm d}\eta \rangle$. Within uncertainties, PYTHIA~8 (tune Monash~2013) and PYTHIA~6 (tune Perugia~2011) quantitatively reproduce the behavior of data, while EPOS~LHC overestimates the value of the exponent. Nevertheless, all models quantitative describe the \pt integrated yields. The use of power-law exponents  facilitates the study of the particle production at high \pt for different collision systems.

Finally, the measurement of the average transverse momentum as a function of event multiplicity at mid-pseudorapidity was presented. The results for the spherocity-integrated class (nearly identical to INEL$\,>0$ pp collisions) at $\sqrt{s}=13$\,TeV are consistent with previous measurements at lower energies. The increase of the average \pt with increasing multiplicity is well captured by  PYTHIA~8 and EPOS~LHC. In order to get a better insight into the particle production mechanisms, the spherocity-integrated sample was separated into different sub-classes characterized by the event structure in the transverse plane. Jet-like and isotropic events were selected based on the spherocity of the events.  Isotropic events are well described by the three models which were considered in this work. Interestingly, PYTHIA~6 reproduces these event classes better than the INEL$\,>0$ sample. For jet-like events, the average \pt is overestimated by PYTHIA~6 and 8 models in the full multiplicity interval reported. However, EPOS~LHC gives the best description of the jet-like event subsample.

The results presented in this paper illustrate the difficulties for the models to describe different observables once they are differentially analyzed  as a function of several variables. The measurements are important to better understand the similarities between heavy-ion and small collision systems, as well as for Monte Carlo tuning purposes.

\newenvironment{acknowledgement}{\relax}{\relax}
\begin{acknowledgement}
\section*{Acknowledgements}

The ALICE Collaboration would like to thank all its engineers and technicians for their invaluable contributions to the construction of the experiment and the CERN accelerator teams for the outstanding performance of the LHC complex.
The ALICE Collaboration gratefully acknowledges the resources and support provided by all Grid centres and the Worldwide LHC Computing Grid (WLCG) collaboration.
The ALICE Collaboration acknowledges the following funding agencies for their support in building and running the ALICE detector:
A. I. Alikhanyan National Science Laboratory (Yerevan Physics Institute) Foundation (ANSL), State Committee of Science and World Federation of Scientists (WFS), Armenia;
Austrian Academy of Sciences, Austrian Science Fund (FWF): [M 2467-N36] and Nationalstiftung f\"{u}r Forschung, Technologie und Entwicklung, Austria;
Ministry of Communications and High Technologies, National Nuclear Research Center, Azerbaijan;
Conselho Nacional de Desenvolvimento Cient\'{\i}fico e Tecnol\'{o}gico (CNPq), Universidade Federal do Rio Grande do Sul (UFRGS), Financiadora de Estudos e Projetos (Finep) and Funda\c{c}\~{a}o de Amparo \`{a} Pesquisa do Estado de S\~{a}o Paulo (FAPESP), Brazil;
Ministry of Science \& Technology of China (MSTC), National Natural Science Foundation of China (NSFC) and Ministry of Education of China (MOEC) , China;
Croatian Science Foundation and Ministry of Science and Education, Croatia;
Centro de Aplicaciones Tecnol\'{o}gicas y Desarrollo Nuclear (CEADEN), Cubaenerg\'{\i}a, Cuba;
Ministry of Education, Youth and Sports of the Czech Republic, Czech Republic;
The Danish Council for Independent Research | Natural Sciences, the Carlsberg Foundation and Danish National Research Foundation (DNRF), Denmark;
Helsinki Institute of Physics (HIP), Finland;
Commissariat \`{a} l'Energie Atomique (CEA), Institut National de Physique Nucl\'{e}aire et de Physique des Particules (IN2P3) and Centre National de la Recherche Scientifique (CNRS) and R\'{e}gion des  Pays de la Loire, France;
Bundesministerium f\"{u}r Bildung und Forschung (BMBF) and GSI Helmholtzzentrum f\"{u}r Schwerionenforschung GmbH, Germany;
General Secretariat for Research and Technology, Ministry of Education, Research and Religions, Greece;
National Research, Development and Innovation Office, Hungary;
Department of Atomic Energy Government of India (DAE), Department of Science and Technology, Government of India (DST), University Grants Commission, Government of India (UGC) and Council of Scientific and Industrial Research (CSIR), India;
Indonesian Institute of Science, Indonesia;
Centro Fermi - Museo Storico della Fisica e Centro Studi e Ricerche Enrico Fermi and Istituto Nazionale di Fisica Nucleare (INFN), Italy;
Institute for Innovative Science and Technology , Nagasaki Institute of Applied Science (IIST), Japan Society for the Promotion of Science (JSPS) KAKENHI and Japanese Ministry of Education, Culture, Sports, Science and Technology (MEXT), Japan;
Consejo Nacional de Ciencia (CONACYT) y Tecnolog\'{i}a, through Fondo de Cooperaci\'{o}n Internacional en Ciencia y Tecnolog\'{i}a (FONCICYT) and Direcci\'{o}n General de Asuntos del Personal Academico (DGAPA), Mexico;
Nederlandse Organisatie voor Wetenschappelijk Onderzoek (NWO), Netherlands;
The Research Council of Norway, Norway;
Commission on Science and Technology for Sustainable Development in the South (COMSATS), Pakistan;
Pontificia Universidad Cat\'{o}lica del Per\'{u}, Peru;
Ministry of Science and Higher Education and National Science Centre, Poland;
Korea Institute of Science and Technology Information and National Research Foundation of Korea (NRF), Republic of Korea;
Ministry of Education and Scientific Research, Institute of Atomic Physics and Ministry of Research and Innovation and Institute of Atomic Physics, Romania;
Joint Institute for Nuclear Research (JINR), Ministry of Education and Science of the Russian Federation, National Research Centre Kurchatov Institute, Russian Science Foundation and Russian Foundation for Basic Research, Russia;
Ministry of Education, Science, Research and Sport of the Slovak Republic, Slovakia;
National Research Foundation of South Africa, South Africa;
Swedish Research Council (VR) and Knut \& Alice Wallenberg Foundation (KAW), Sweden;
European Organization for Nuclear Research, Switzerland;
National Science and Technology Development Agency (NSDTA), Suranaree University of Technology (SUT) and Office of the Higher Education Commission under NRU project of Thailand, Thailand;
Turkish Atomic Energy Agency (TAEK), Turkey;
National Academy of  Sciences of Ukraine, Ukraine;
Science and Technology Facilities Council (STFC), United Kingdom;
National Science Foundation of the United States of America (NSF) and United States Department of Energy, Office of Nuclear Physics (DOE NP), United States of America.    
\end{acknowledgement}

\bibliographystyle{utphys}   
\bibliography{biblio}


\newpage
\appendix
\section{The ALICE Collaboration}
\label{app:collab}

\begingroup
\small
\begin{flushleft}
S.~Acharya\Irefn{org141}\And 
D.~Adamov\'{a}\Irefn{org93}\And 
S.P.~Adhya\Irefn{org141}\And 
A.~Adler\Irefn{org74}\And 
J.~Adolfsson\Irefn{org80}\And 
M.M.~Aggarwal\Irefn{org98}\And 
G.~Aglieri Rinella\Irefn{org34}\And 
M.~Agnello\Irefn{org31}\And 
N.~Agrawal\Irefn{org10}\And 
Z.~Ahammed\Irefn{org141}
% ...
\And
\renewcommand\labelenumi{\textsuperscript{\theenumi}~}

\section*{Affiliation notes}
\renewcommand\theenumi{\roman{enumi}}
\begin{Authlist}
\item \Adef{org*}Deceased
\item \Adef{orgI}Dipartimento DET del Politecnico di Torino, Turin, Italy
\item \Adef{orgII}M.V. Lomonosov Moscow State University, D.V. Skobeltsyn Institute of Nuclear, Physics, Moscow, Russia
\item \Adef{orgIII}Department of Applied Physics, Aligarh Muslim University, Aligarh, India
\item \Adef{orgIV}Institute of Theoretical Physics, University of Wroclaw, Poland
\end{Authlist}

\section*{Collaboration Institutes}
\renewcommand\theenumi{\arabic{enumi}~}
\begin{Authlist}
\item \Idef{org1}A.I. Alikhanyan National Science Laboratory (Yerevan Physics Institute) Foundation, Yerevan, Armenia
\item \Idef{org2}Bogolyubov Institute for Theoretical Physics, National Academy of Sciences of Ukraine, Kiev, Ukraine
\item \Idef{org3}Bose Institute, Department of Physics  and Centre for Astroparticle Physics and Space Science (CAPSS), Kolkata, India
\item \Idef{org4}Budker Institute for Nuclear Physics, Novosibirsk, Russia
\item \Idef{org5}California Polytechnic State University, San Luis Obispo, California, United States
\item \Idef{org6}Central China Normal University, Wuhan, China
\item \Idef{org7}Centre de Calcul de l'IN2P3, Villeurbanne, Lyon, France
\item \Idef{org8}Centro de Aplicaciones Tecnol\'{o}gicas y Desarrollo Nuclear (CEADEN), Havana, Cuba
\item \Idef{org9}Centro de Investigaci\'{o}n y de Estudios Avanzados (CINVESTAV), Mexico City and M\'{e}rida, Mexico
\item \Idef{org10}Centro Fermi - Museo Storico della Fisica e Centro Studi e Ricerche ``Enrico Fermi', Rome, Italy
\item \Idef{org11}Chicago State University, Chicago, Illinois, United States
\item \Idef{org12}China Institute of Atomic Energy, Beijing, China
\item \Idef{org13}Chonbuk National University, Jeonju, Republic of Korea
\item \Idef{org14}Comenius University Bratislava, Faculty of Mathematics, Physics and Informatics, Bratislava, Slovakia
\item \Idef{org15}COMSATS University Islamabad, Islamabad, Pakistan
\item \Idef{org16}Creighton University, Omaha, Nebraska, United States
\item \Idef{org17}Department of Physics, Aligarh Muslim University, Aligarh, India
\item \Idef{org18}Department of Physics, Pusan National University, Pusan, Republic of Korea
\item \Idef{org19}Department of Physics, Sejong University, Seoul, Republic of Korea
\item \Idef{org20}Department of Physics, University of California, Berkeley, California, United States
\item \Idef{org21}Department of Physics, University of Oslo, Oslo, Norway
\item \Idef{org22}Department of Physics and Technology, University of Bergen, Bergen, Norway
\item \Idef{org23}Dipartimento di Fisica dell'Universit\`{a} 'La Sapienza' and Sezione INFN, Rome, Italy
\item \Idef{org24}Dipartimento di Fisica dell'Universit\`{a} and Sezione INFN, Cagliari, Italy
\item \Idef{org25}Dipartimento di Fisica dell'Universit\`{a} and Sezione INFN, Trieste, Italy
\item \Idef{org26}Dipartimento di Fisica dell'Universit\`{a} and Sezione INFN, Turin, Italy
\item \Idef{org27}Dipartimento di Fisica e Astronomia dell'Universit\`{a} and Sezione INFN, Bologna, Italy
\item \Idef{org28}Dipartimento di Fisica e Astronomia dell'Universit\`{a} and Sezione INFN, Catania, Italy
\item \Idef{org29}Dipartimento di Fisica e Astronomia dell'Universit\`{a} and Sezione INFN, Padova, Italy
\item \Idef{org30}Dipartimento di Fisica `E.R.~Caianiello' dell'Universit\`{a} and Gruppo Collegato INFN, Salerno, Italy
\item \Idef{org31}Dipartimento DISAT del Politecnico and Sezione INFN, Turin, Italy
\item \Idef{org32}Dipartimento di Scienze e Innovazione Tecnologica dell'Universit\`{a} del Piemonte Orientale and INFN Sezione di Torino, Alessandria, Italy
\item \Idef{org33}Dipartimento Interateneo di Fisica `M.~Merlin' and Sezione INFN, Bari, Italy
\item \Idef{org34}European Organization for Nuclear Research (CERN), Geneva, Switzerland
\item \Idef{org35}Faculty of Electrical Engineering, Mechanical Engineering and Naval Architecture, University of Split, Split, Croatia
\item \Idef{org36}Faculty of Engineering and Science, Western Norway University of Applied Sciences, Bergen, Norway
\item \Idef{org37}Faculty of Nuclear Sciences and Physical Engineering, Czech Technical University in Prague, Prague, Czech Republic
\item \Idef{org38}Faculty of Science, P.J.~\v{S}af\'{a}rik University, Ko\v{s}ice, Slovakia
\item \Idef{org39}Frankfurt Institute for Advanced Studies, Johann Wolfgang Goethe-Universit\"{a}t Frankfurt, Frankfurt, Germany
\item \Idef{org40}Gangneung-Wonju National University, Gangneung, Republic of Korea
\item \Idef{org41}Gauhati University, Department of Physics, Guwahati, India
\item \Idef{org42}Helmholtz-Institut f\"{u}r Strahlen- und Kernphysik, Rheinische Friedrich-Wilhelms-Universit\"{a}t Bonn, Bonn, Germany
\item \Idef{org43}Helsinki Institute of Physics (HIP), Helsinki, Finland
\item \Idef{org44}High Energy Physics Group,  Universidad Aut\'{o}noma de Puebla, Puebla, Mexico
\item \Idef{org45}Hiroshima University, Hiroshima, Japan
\item \Idef{org46}Hochschule Worms, Zentrum  f\"{u}r Technologietransfer und Telekommunikation (ZTT), Worms, Germany
\item \Idef{org47}Horia Hulubei National Institute of Physics and Nuclear Engineering, Bucharest, Romania
\item \Idef{org48}Indian Institute of Technology Bombay (IIT), Mumbai, India
\item \Idef{org49}Indian Institute of Technology Indore, Indore, India
\item \Idef{org50}Indonesian Institute of Sciences, Jakarta, Indonesia
\item \Idef{org51}INFN, Laboratori Nazionali di Frascati, Frascati, Italy
\item \Idef{org52}INFN, Sezione di Bari, Bari, Italy
\item \Idef{org53}INFN, Sezione di Bologna, Bologna, Italy
\item \Idef{org54}INFN, Sezione di Cagliari, Cagliari, Italy
\item \Idef{org55}INFN, Sezione di Catania, Catania, Italy
\item \Idef{org56}INFN, Sezione di Padova, Padova, Italy
\item \Idef{org57}INFN, Sezione di Roma, Rome, Italy
\item \Idef{org58}INFN, Sezione di Torino, Turin, Italy
\item \Idef{org59}INFN, Sezione di Trieste, Trieste, Italy
\item \Idef{org60}Inha University, Incheon, Republic of Korea
\item \Idef{org61}Institut de Physique Nucl\'{e}aire d'Orsay (IPNO), Institut National de Physique Nucl\'{e}aire et de Physique des Particules (IN2P3/CNRS), Universit\'{e} de Paris-Sud, Universit\'{e} Paris-Saclay, Orsay, France
\item \Idef{org62}Institute for Nuclear Research, Academy of Sciences, Moscow, Russia
\item \Idef{org63}Institute for Subatomic Physics, Utrecht University/Nikhef, Utrecht, Netherlands
\item \Idef{org64}Institute for Theoretical and Experimental Physics, Moscow, Russia
\item \Idef{org65}Institute of Experimental Physics, Slovak Academy of Sciences, Ko\v{s}ice, Slovakia
\item \Idef{org66}Institute of Physics, Homi Bhabha National Institute, Bhubaneswar, India
\item \Idef{org67}Institute of Physics of the Czech Academy of Sciences, Prague, Czech Republic
\item \Idef{org68}Institute of Space Science (ISS), Bucharest, Romania
\item \Idef{org69}Institut f\"{u}r Kernphysik, Johann Wolfgang Goethe-Universit\"{a}t Frankfurt, Frankfurt, Germany
\item \Idef{org70}Instituto de Ciencias Nucleares, Universidad Nacional Aut\'{o}noma de M\'{e}xico, Mexico City, Mexico
\item \Idef{org71}Instituto de F\'{i}sica, Universidade Federal do Rio Grande do Sul (UFRGS), Porto Alegre, Brazil
\item \Idef{org72}Instituto de F\'{\i}sica, Universidad Nacional Aut\'{o}noma de M\'{e}xico, Mexico City, Mexico
\item \Idef{org73}iThemba LABS, National Research Foundation, Somerset West, South Africa
\item \Idef{org74}Johann-Wolfgang-Goethe Universit\"{a}t Frankfurt Institut f\"{u}r Informatik, Fachbereich Informatik und Mathematik, Frankfurt, Germany
\item \Idef{org75}Joint Institute for Nuclear Research (JINR), Dubna, Russia
\item \Idef{org76}Korea Institute of Science and Technology Information, Daejeon, Republic of Korea
\item \Idef{org77}KTO Karatay University, Konya, Turkey
\item \Idef{org78}Laboratoire de Physique Subatomique et de Cosmologie, Universit\'{e} Grenoble-Alpes, CNRS-IN2P3, Grenoble, France
\item \Idef{org79}Lawrence Berkeley National Laboratory, Berkeley, California, United States
\item \Idef{org80}Lund University Department of Physics, Division of Particle Physics, Lund, Sweden
\item \Idef{org81}Nagasaki Institute of Applied Science, Nagasaki, Japan
\item \Idef{org82}Nara Women{'}s University (NWU), Nara, Japan
\item \Idef{org83}National and Kapodistrian University of Athens, School of Science, Department of Physics , Athens, Greece
\item \Idef{org84}National Centre for Nuclear Research, Warsaw, Poland
\item \Idef{org85}National Institute of Science Education and Research, Homi Bhabha National Institute, Jatni, India
\item \Idef{org86}National Nuclear Research Center, Baku, Azerbaijan
\item \Idef{org87}National Research Centre Kurchatov Institute, Moscow, Russia
\item \Idef{org88}Niels Bohr Institute, University of Copenhagen, Copenhagen, Denmark
\item \Idef{org89}Nikhef, National institute for subatomic physics, Amsterdam, Netherlands
\item \Idef{org90}NRC Kurchatov Institute IHEP, Protvino, Russia
\item \Idef{org91}NRNU Moscow Engineering Physics Institute, Moscow, Russia
\item \Idef{org92}Nuclear Physics Group, STFC Daresbury Laboratory, Daresbury, United Kingdom
\item \Idef{org93}Nuclear Physics Institute of the Czech Academy of Sciences, \v{R}e\v{z} u Prahy, Czech Republic
\item \Idef{org94}Oak Ridge National Laboratory, Oak Ridge, Tennessee, United States
\item \Idef{org95}Ohio State University, Columbus, Ohio, United States
\item \Idef{org96}Petersburg Nuclear Physics Institute, Gatchina, Russia
\item \Idef{org97}Physics department, Faculty of science, University of Zagreb, Zagreb, Croatia
\item \Idef{org98}Physics Department, Panjab University, Chandigarh, India
\item \Idef{org99}Physics Department, University of Jammu, Jammu, India
\item \Idef{org100}Physics Department, University of Rajasthan, Jaipur, India
\item \Idef{org101}Physikalisches Institut, Eberhard-Karls-Universit\"{a}t T\"{u}bingen, T\"{u}bingen, Germany
\item \Idef{org102}Physikalisches Institut, Ruprecht-Karls-Universit\"{a}t Heidelberg, Heidelberg, Germany
\item \Idef{org103}Physik Department, Technische Universit\"{a}t M\"{u}nchen, Munich, Germany
\item \Idef{org104}Politecnico di Bari, Bari, Italy
\item \Idef{org105}Research Division and ExtreMe Matter Institute EMMI, GSI Helmholtzzentrum f\"ur Schwerionenforschung GmbH, Darmstadt, Germany
\item \Idef{org106}Rudjer Bo\v{s}kovi\'{c} Institute, Zagreb, Croatia
\item \Idef{org107}Russian Federal Nuclear Center (VNIIEF), Sarov, Russia
\item \Idef{org108}Saha Institute of Nuclear Physics, Homi Bhabha National Institute, Kolkata, India
\item \Idef{org109}School of Physics and Astronomy, University of Birmingham, Birmingham, United Kingdom
\item \Idef{org110}Secci\'{o}n F\'{\i}sica, Departamento de Ciencias, Pontificia Universidad Cat\'{o}lica del Per\'{u}, Lima, Peru
\item \Idef{org111}Shanghai Institute of Applied Physics, Shanghai, China
\item \Idef{org112}St. Petersburg State University, St. Petersburg, Russia
\item \Idef{org113}Stefan Meyer Institut f\"{u}r Subatomare Physik (SMI), Vienna, Austria
\item \Idef{org114}SUBATECH, IMT Atlantique, Universit\'{e} de Nantes, CNRS-IN2P3, Nantes, France
\item \Idef{org115}Suranaree University of Technology, Nakhon Ratchasima, Thailand
\item \Idef{org116}Technical University of Ko\v{s}ice, Ko\v{s}ice, Slovakia
\item \Idef{org117}Technische Universit\"{a}t M\"{u}nchen, Excellence Cluster 'Universe', Munich, Germany
\item \Idef{org118}The Henryk Niewodniczanski Institute of Nuclear Physics, Polish Academy of Sciences, Cracow, Poland
\item \Idef{org119}The University of Texas at Austin, Austin, Texas, United States
\item \Idef{org120}Universidad Aut\'{o}noma de Sinaloa, Culiac\'{a}n, Mexico
\item \Idef{org121}Universidade de S\~{a}o Paulo (USP), S\~{a}o Paulo, Brazil
\item \Idef{org122}Universidade Estadual de Campinas (UNICAMP), Campinas, Brazil
\item \Idef{org123}Universidade Federal do ABC, Santo Andre, Brazil
\item \Idef{org124}University College of Southeast Norway, Tonsberg, Norway
\item \Idef{org125}University of Cape Town, Cape Town, South Africa
\item \Idef{org126}University of Houston, Houston, Texas, United States
\item \Idef{org127}University of Jyv\"{a}skyl\"{a}, Jyv\"{a}skyl\"{a}, Finland
\item \Idef{org128}University of Liverpool, Liverpool, United Kingdom
\item \Idef{org129}University of Science and Techonology of China, Hefei, China
\item \Idef{org130}University of Tennessee, Knoxville, Tennessee, United States
\item \Idef{org131}University of the Witwatersrand, Johannesburg, South Africa
\item \Idef{org132}University of Tokyo, Tokyo, Japan
\item \Idef{org133}University of Tsukuba, Tsukuba, Japan
\item \Idef{org134}Universit\'{e} Clermont Auvergne, CNRS/IN2P3, LPC, Clermont-Ferrand, France
\item \Idef{org135}Universit\'{e} de Lyon, Universit\'{e} Lyon 1, CNRS/IN2P3, IPN-Lyon, Villeurbanne, Lyon, France
\item \Idef{org136}Universit\'{e} de Strasbourg, CNRS, IPHC UMR 7178, F-67000 Strasbourg, France, Strasbourg, France
\item \Idef{org137}Universit\'{e} Paris-Saclay Centre d'Etudes de Saclay (CEA), IRFU, D\'{e}partment de Physique Nucl\'{e}aire (DPhN), Saclay, France
\item \Idef{org138}Universit\`{a} degli Studi di Foggia, Foggia, Italy
\item \Idef{org139}Universit\`{a} degli Studi di Pavia, Pavia, Italy
\item \Idef{org140}Universit\`{a} di Brescia, Brescia, Italy
\item \Idef{org141}Variable Energy Cyclotron Centre, Homi Bhabha National Institute, Kolkata, India
\item \Idef{org142}Warsaw University of Technology, Warsaw, Poland
\item \Idef{org143}Wayne State University, Detroit, Michigan, United States
\item \Idef{org144}Westf\"{a}lische Wilhelms-Universit\"{a}t M\"{u}nster, Institut f\"{u}r Kernphysik, M\"{u}nster, Germany
\item \Idef{org145}Wigner Research Centre for Physics, Hungarian Academy of Sciences, Budapest, Hungary
\item \Idef{org146}Yale University, New Haven, Connecticut, United States
\item \Idef{org147}Yonsei University, Seoul, Republic of Korea
\end{Authlist}
\endgroup
\end{document}